\documentclass[conference]{IEEEtran}
\IEEEoverridecommandlockouts
\usepackage{cite}
\usepackage{amsmath,amssymb,amsfonts}
\usepackage{graphicx}
\usepackage{textcomp}
\usepackage{xcolor}
\usepackage{algorithm}
\usepackage{algorithmicx}
\usepackage{algpseudocode}
\usepackage{multirow}
\usepackage{subfigure}
\usepackage{float}
\usepackage{makecell}
\usepackage{color}
\usepackage{hyperref}

\newtheorem{definition}{Definition}

\floatname{algorithm}{Algorithm}

\def\BibTeX{{\rm B\kern-.05em{\sc i\kern-.025em b}\kern-.08em
    T\kern-.1667em\lower.7ex\hbox{E}\kern-.125emX}}
\begin{document}

\title{A Unified Replay-based Continuous Learning Framework for Spatio-Temporal Prediction\\ on Streaming Data}

\author{\IEEEauthorblockN{Hao Miao$^1$, Yan Zhao$^{1,*}$, Chenjuan Guo$^{1,2,*}$, Bin Yang$^{1, 2}$, Kai Zheng$^3$, Feiteng Huang$^4$, Jiandong Xie$^4$, \\Christian S. Jensen$^1$}
\IEEEauthorblockA{$^{1}$Aalborg University, Denmark\\
$^{2}$East China Normal University, China\\
$^{3}$University of Electronic Science and Technology of China, China\\
$^{4}$Huawei Cloud Database Innovation Lab, China\\
\{haom, yanz, cguo, byang\}@cs.aau.dk, zhengkai@uestc.edu.cn, \{huangfeiteng, xiejiandong\}@huawei.com, csj@cs.aau.dk
}
% \and
% \and
% \and
% \and
% \and
}

\maketitle

\renewcommand{\thefootnote}{\fnsymbol{footnote}}
\footnotetext[1]{Corresponding authors.}

\begin{abstract}
 The widespread deployment of wireless and mobile devices results in a proliferation of spatio-temporal data that is used in applications, e.g., traffic prediction, human mobility mining, and air quality prediction, where spatio-temporal prediction is often essential to enable safety, predictability, or reliability. Many recent proposals that target deep learning for spatio-temporal prediction suffer from so-called catastrophic forgetting, where previously learned knowledge is entirely forgotten when new data arrives. Such proposals may experience deteriorating prediction performance when applied in settings where data streams into the system. To enable spatio-temporal prediction on streaming data, we propose a unified replay-based continuous learning framework. The framework includes a replay buffer of previously learned samples that are fused with training data using a spatio-temporal mixup mechanism in order to preserve historical knowledge effectively, thus avoiding catastrophic forgetting. To enable holistic representation preservation, the framework also integrates a general spatio-temporal autoencoder with a carefully designed spatio-temporal simple siamese (STSimSiam) network that aims to ensure prediction accuracy and avoid holistic feature loss by means of mutual information maximization. The framework further encompasses five spatio-temporal data augmentation methods to enhance the performance of STSimSiam. Extensive experiments on real data offer insight into the effectiveness of the proposed framework.
\end{abstract}

\begin{IEEEkeywords}
Spatio-Temporal Prediction, Continuous Learning, Streaming Data
\end{IEEEkeywords}

\section{Introduction}
The continued digitization of societal processes and the accompanying deployment of sensing technologies generate increasingly massive amounts of spatio-temporal data. For example, populations of in-road sensors provide data that captures traffic flow in multiple locations across time. Further, applications increasingly ingest large amounts of spatio-temporal data that is being generated continuously, which is called streaming spatio-temporal data. The data from in-road sensors is an example of such data. 
%To enable value creation from such data, effective spatio-temporal prediction techniques are needed to \iffalse that can \fi predict the unknown data or system states by considering correlations among both the spatial and temporal dimensions, which have drawn considerable attention in both academia and industry~\cite{zhang2017deep, jin2022selective, feng2018deepmove, yi2018deep, miao2022mba}. 
%Spatio-temporal prediction plays a vital role in various urban computing applications for the construction of smart cities. For instance, precise traffic prediction can facilitate more effective traffic management and improve vehicle routing in intelligent transportation systems \cite{tong2018unified}.
%Spatio-temporal prediction plays a vital role in various urban computing applications for the construction of smart cities, such as traffic prediction~\cite{zhang2017deep, jin2022selective}, human mobility mining~\cite{feng2018deepmove}, and air quality prediction~\cite{yi2018deep}.
%For instance, precise traffic prediction can facilitate more effective traffic management and improve vehicle routing in intelligent transportation systems \cite{tong2018unified}.
%\HAO{In this work, we study the problem of spatio-temporal prediction aiming at learning .} 

In this study, we focus on the novel problem of prediction on streaming spatio-temporal data, where the general goal is to learn a model from streaming spatio-temporal data with spatial and temporal correlations while preserving learned historical knowledge and capturing spatio-temporal patterns to accurately predict future spatio-temporal observations.

Many spatio-temporal prediction applications, e.g., traffic flow prediction~\cite{zhang2017deep, miao2022mba, ji2022stden}, traffic speed prediction~\cite{li2018diffusion, fang2021spatial} and on-demand service prediction~\cite{ yao2019revisiting}, exist along with accompanying, well-customized prediction models. These models use a variety of means to predict spatio-temporal data, including approaches based on traditional statistics~\cite{smith1997traffic, shekhar2007adaptive} and convolutional~\cite{wu2019graph, wu2020connecting, razvanicde2021} and recurrent~\cite{li2018diffusion, bai2020adaptive} neural networks. However, existing models are trained statically and fail to handle streaming data. Static models are often trained once to fit a particular dataset and then make predictions that aim to fit another dataset with the underlying assumption that the two datasets follow the same distribution. However, concept drift, meaning that data distributions change over time, occurs often in streaming spatio-temporal data. Thus, direct application of static models to streaming data may cause remarkable performance degradation~\cite{serra2018overcoming}.

As a result, we need a new kind of continuous learning (CL) model that can adapt continually to the data it sees and can keep on learning spatio-temporal prediction tasks over time. However, it is non-trivial to develop this kind of model, due to the following challenges.

\emph{Challenge I: catastrophic forgetting.} It is challenging to alleviate catastrophic forgetting in CL for spatio-temporal prediction. Catastrophic forgetting is the tendency to abruptly forget previously learned knowledge, which occurs when static models are simply retrained using newly arrived data~\cite{serra2018overcoming}. When a model is retrained continuously based on incoming data, the distributions of which are different (due to concept drift), prediction performance on preceding tasks deteriorates~\cite{hofmanninger2020dynamic}. This is because the model is always retrained on new data obtained during one period and is then used to predict for another period. When the data arrives continuously, the knowledge learned by the model keeps drifting. Although many efforts have been made to use CL techniques to address catastrophic forgetting in computer vision~\cite{kirkpatrick2017overcoming, NeurIPS2019_9357} and natural language processing~\cite{sun2019lamol}, these techniques cannot be applied to spatio-temporal prediction directly due to the unique characteristics of spatio-temporal data. Specifically, no CL model exists that can effectively capture spatial dependencies or temporal correlations in streaming spatio-temporal data.

%\emph{Challenge I: catastrophic forgetting.} It is challenging to alleviate catastrophic forgetting in CL for spatio-temporal prediction. Catastrophic forgetting is the tendency to abruptly forget previously learned knowledge if the static models are simply retrained using the new arrived data~\cite{mccloskey1989catastrophic,serra2018overcoming}. \iffalse which is a major undesired issue affecting CL \fi When a model is continuously updated on new-coming data, the distributions of which are different (i.e., concept drift), the model performance deteriorates on the preceding task~\cite{hofmanninger2020dynamic} in terms of prediction accuracy. This is because that the model is always retrained with new data obtained in a period and is used to predict for another period. When the data keeps arriving, the knowledge learned by the model keeps drifting. Although many efforts have been made to use CL techniques to address the catastrophic forgetting issue in computer vision~\cite{kirkpatrick2017overcoming, NeurIPS2019_9357} and natural language processing~\cite{sun2019lamol, ke2021achieving}, these techniques cannot be applied to spatio-temporal prediction directly due to the unique spatial and temporal patterns. Specifically, there is no tailor-made CL model that can effectively capture spatial dependencies and/or temporal correlations of spatio-temporal data.
%, and multivariate time series tasks~\cite{gupta2021continual}

\emph{Challenge II: diversity of spatio-temporal data and prediction applications.}  It remains a key problem to discover the commonalities of diverse spatio-temporal data and prediction applications. Although various well-customized static models exist, it is time-consuming and unrealistic to convert each such model to its continuous version since there are different specific settings (e.g., network architectures, datasets, and objective functions) to consider for each spatio-temporal prediction model. It is thus highly desirable but also non-trivial to achieve a model that is accurate and efficient across many streaming spatio-temporal prediction applications.

\emph{Challenge III: holistic feature preservation.} It is challenging for existing models to learn holistic features for spatio-temporal prediction on streaming data. Specifically, holistic features preserve semantic similarities across multiple time periods. In continuous spatio-temporal prediction, preserving previously learned semantic features may facilitate future prediction. For example, traffic patterns learned during off-peak hours on previous weekdays may be helpful for prediction of subsequent weekdays. Most existing CL models focus on learning discriminative features for a current task, while ignoring previously learned features that may be useful for future tasks~\cite{guo2022online}, which often lead to unsatisfactory results for future prediction and thus affect continuous spatio-temporal prediction adversely.

This study addresses the above challenges by providing a Unified Replay-based Continuous Learning (URCL) framework for spatio-temporal prediction on streaming data. URCL encompasses three main modules: data integration, spatio-temporal continuous representation learning (STCRL), and spatio-temporal prediction. To alleviate catastrophic forgetting (Challenge I), we propose a spatio-temporal mixup (STMixup) mechanism to fuse current sampled spatio-temporal observations with selected samples from a replay buffer that stores a subset of previously learned observations. We also propose a ranking-based maximally interfered retrieval sampling strategy to select representative samples from the buffer.

To support diverse spatio-temporal data and prediction applications (Challenge II), we discover the commonalities of existing methods that are typically based on an autoencoder architecture. We propose a novel spatio-temporal prediction network including a spatio-temporal encoder (STEncoder) and a spatio-temporal decoder (STDecoder) to capture complex spatio-temporal correlations to enable accuracy.

To address the issue of holistic-feature preservation (Challenge III), we propose a spatio-temporal simple siamese (STSimSiam) network to avoid holistic feature loss in the STCRL module. Specifically, the STSimSiam
network contains two STEncoders and a projection multi-layer perceptron (MLP) head, where the STEncoders are shared with that of the spatio-temporal prediction network. We first use mutual
information maximization to ensure holistic feature preservation in streaming spatio-temporal prediction. In addition, considering that data augmentation can help the model learn more effective representations~\cite{chen2021exploring, zbontar2021barlow}, we provide five spatio-temporal data augmentation methods based on an exploration of the exclusive characteristics of spatio-temporal data, thus achieving effective holistic spatio-temporal feature learning.

The major contributions are summarized as follows. 

\begin{itemize}
    \item To the best of our knowledge, this is the first study to systematically investigate continuous learning for spatio-temporal prediction on streaming data. We propose a unified replay-based continuous learning framework entitled URCL for spatio-temporal prediction on streaming data, where a replay buffer and a spatio-temporal mixup mechanism are designed to alleviate catastrophic forgetting.
    %To the best of our knowledge, this is the first study to systematically investigate continuous learning in spatio-temporal prediction on streaming data. We propose a unified replay-based continuous learning framework entitled URCL for spatio-temporal prediction on streaming data, where a replay buffer and a spatio-temporal mixup mechanism are designed to alleviate catastrophic forgetting.
    \item To contend with the diversity of spatio-temporal data and prediction applications, we design a spatio-temporal autoencoder including an STEncoder and an STDecoder for effective spatio-temporal feature learning and prediction.
    \item A novel STSimSiam network is designed to make learned features in latent spaces more holistic, by applying mutual information maximization to preserve features for continuous spatio-temporal prediction. We also provide five data augmentation methods by considering unique spatiotemporal properties.
    %A novel STSimSiam network is designed to make the learned features in latent spaces more holistic, where mutual information maximization is applied to preserve features for continuous spatio-temporal prediction. We also give five data argumentation methods by considering unique spatio-temporal properties.
    %\item A self-supervised learning loss and a prediction loss are integrated into the continuous learning framework to make the learned features in latent spaces more holistic, where mutual information maximization is firstly applied to preserve features for continuous spatio-temporal prediction. We also give five data argumentation methods by considering unique spatio-temporal properties.
    \item We report on experiments using real datasets, offering evidence of the effectiveness of the proposed URCL.
\end{itemize}

%\begin{itemize}
%    \item To the best of our knowledge, this is the first study to systematically investigate replay-based continuous learning spatio-temporal prediction on streaming data.
%    \item We propose a unified replay-based continuous learning framework named URCL to address this problem via joint training. A self-supervised learning loss (i.e., GraphCL) and forecasting loss are integrated into the continuous learning framework to make the learned features in latent spaces are more holistic.
%    \item STMixup is designed to interpolate between current data and previous instances to alleviate catastrophic forgetting. In addition, we propose five types of data argumentation methods as auxiliary of spatio-temporal self-supervised learning with the consideration of unique spatio-temporal properties.
%\end{itemize}

The remainder of this paper is organized as follows. Section \ref{RelatedWork} surveys the related work, and Section \ref{problem} introduces preliminary concepts and the streaming spatio-temporal prediction problem. We then present the URCL framework in Section \ref{method}, followed by the experimental results in Section \ref{experiment}. Section \ref{conclusion} concludes the paper.

\section{Related work}
\label{RelatedWork}
%This work is highly relevant to the research topics of spatio-temporal data forecasting, and continual learning. We will review related works from the two aspects.
We briefly review prior studies on spatio-temporal data prediction and continuous learning.
\subsection{Spatio-Temporal Data Prediction}
Spatio-temporal data prediction attracts increasing interest due to the increasing availability of spatio-temporal data and rich applications, such as traffic prediction~\cite{wang2020deep, zhang2017deep, ji2023spatio, zhou2023Predicting}, precipitation prediction~\cite{shi2015convolutional}, and air quality prediction~\cite{yi2018deep}. 
%In this work, we select traffic prediction as an example of spatio-temporal data prediction. 
Traditional spatio-temporal prediction models are mostly based on statistical models~\cite{shekhar2007adaptive, lippi2013short}. However, the statistical models cannot capture complex spatial and temporal correlations of spatio-temporal data effectively due to their limited learning capacity.

With the advance of deep learning techniques, various deep learning based methods address spatio-temporal data prediction~\cite{kieu2022anomaly, li2018diffusion, yanwww2022, kieu2022robust, mo2020towards, chen2021daemon, chen2023adversarial, DBLP:journals/pacmmod/0002Z0KGJ23, DBLP:journals/pacmmod/Wu0ZG0J23, wupvldb}, and outperform
%have achieved much better performance better than 
traditional statistical models. One line of study~\cite{miao2022mba, zhang2017deep} treats the spatio-temporal data of an entire city as images and applies convolutional neural network (CNN) to extract spatial correlations. %Zhang et al.~\cite{zhang2017deep} propose a deep-learning-based approach, called ST-ResNet, which is composed of stacked CNN layers and predicts the inflow and the outflow of crowds in regions of a city. Yao et al.~\cite{yao2019revisiting} design a Spatial-Temporal Dynamic Network combining CNNs and recurrent neural networks (RNNs) for grid-based traffic prediction. 
Another line of study~\cite{wu2019graph, li2021spatial, li2018diffusion,bai2020adaptive, DBLP:conf/kdd/ZhouHYWWZLW23, yunyaovldb24, kaivldb24} employs graph neural network to perform spatio-temporal prediction by modeling global spatial dependencies and local spatial correlations effectively. %Li et al.~\cite{li2018diffusion} propose a Diffusion Convolutional Recurrent Neural Network (DCRNN) for sensor-based traffic prediction on a directed road graph. Bai et al.~\cite{bai2020adaptive} give an Adaptive Graph Convolutional Recurrent Network (AGCRN) to jointly capture fine-grained spatial and temporal correlations. 
However, these methods cannot support stream setting and suffer from catastrophic forgetting, which makes continuous or lifelong learning difficult for spatio-temporal prediction.
%Even though the above mentioned methods are able to achieve excellent performance, there is no existing work considering spatio-temporal prediction in a continual setting.

\subsection{Continuous Learning}
Continuous learning, which is also called lifelong or incremental learning, learns a sequence of tasks incrementally with knowledge transfer and without catastrophic forgetting~\cite{chen2018lifelong}. The goal of continuous learning is to extend knowledge acquired gradually from an infinite data stream and use it for future learning.%~\cite{delange2021continual}. 

In the early stage, continuous learning studies target the domain of object recognition~\cite{thrun1995learning, ruvolo2013ella}. Thrun et al.~\cite{thrun1995learning} propose several lifelong learning algorithms encompassing memory-based and neural network-based approaches. Ruvolo et al.~\cite{ruvolo2013ella} design an Efficient Lifelong Learning Algorithm for online multi-task learning.% Mitchell et al.~\cite{mitchell2018never} reformulate a never-ending language paradigm for machine learning and design a never-ending language learner to enable intelligent agents to learn many types of knowledge continuously.

With the rapid development of deep learning techniques, focus is on deep learning based continuous learning methods~\cite{kirkpatrick2017overcoming, NeurIPS2019_9357}, which can be divided into replay-based, regularization-based, and architecture-based methods. Specifically, replay-based methods~\cite{NeurIPS2019_9357} usually adopt an explicit buffer to store a subset of training samples or learn a generator. %For instance, Rahaf et al.~\cite{NeurIPS2019_9357} propose a maximization-based interfered retrieval method to select samples from a buffer to enhance continuous learning. 
Regularization-based methods include a regularization term in the loss function~\cite{kirkpatrick2017overcoming, aljundi2018memory}. %For example, an Elastic Weight Consolidation (EWC)~\cite{kirkpatrick2017overcoming} method is proposed to remember old tasks by selectively slowing down learning of the weights that are important for those tasks.
%overcome catastrophic forgetting by selectively slowing down learning on the weights important for old tasks.
In architecture-based methods~\cite{serra2018overcoming}, a different sub-network is dedicated to each incremental learning task. %Serra et al.~\cite{wortsman2020supermasks} propose a Superposition (SupSup) model with Supermasks, which adopts a randomly initialized, fixed base network and finds a sub-network (called supermask) for each task. SupSup is capable of learning thousands of tasks sequentially without catastrophic forgetting. 
However, most of the above methods are designed for computer vision and natural language processing, and cannot be applied to spatio-temporal data prediction directly due to complex spatio-temporal patterns and unique spatio-temporal semantics. 

Although previous studies~\cite{Chen2021IJCAI, xiao2022streaming} predict streaming traffic flow, their problem settings differ substantially from ours that focusing on node incremental learning where the number of traffic sensors varies across time. In our work, the number of sensors does not vary, while the instances vary across time.

%while ignoring developing a continuous learning framework for spatio-temporal data prediction.

\section{Problem Statement}
\label{problem}
%In this section, we will first give some basic notations to help us state the studied problem. Then, a formal problem definition will be given.
We proceed to present necessary preliminaries and then define the problem addressed.

Advances in hardware and wireless network technologies have resulted in multi-functional sensor devices~\cite{tubaishat2003sensor}. 
This development enables systems, called sensor networks, consisting of tiny sensor nodes spread across large geographical areas, recording streaming spatio-temporal data.

\begin{definition}[Sensor Network]
%\textbf{Road sensor network} 
A sensor network is denoted by a graph $G = (V, E)$, where $V$ is a sensor node set, and $E$ is an edge set.
Each node $v_i \in V$ represents a sensor, and 
%deployed on a road segment. 
each edge $e_{i,j} \in E$ indicates connectivity between sensors $v_i$ and $v_j$. % that are deployed on two neighbor road links.
%We denote a road sensor network as a graph $G = \{V, E\}$, where $|V| = N$ is the vertices set, $N$ denotes the number of vertices, $E$ is the edge set. Each node $v_i \in V$ represents a traffic sensor deployed on a road segment. Edge $e_{i,j} \in E$ indicates the connectivity between two road sensors $v_i$ and $v_j$ that are deployed on two neighbor road links.
\end{definition}

\begin{definition}[Spatio-temporal Observation]
%\textbf{Network signal observations}
%The network signal observations collected by all the sensors over $T$ time intervals are defined as $\mathcal{X} = \left\langle X^1, X^2,  \cdots, X^T\right\rangle$. 
Given a sensor network $G$, a spatio-temporal observation collected by all sensors at time slot $t$ (e.g., 9:00 a.m.--9:15 a.m.) with a sampling interval $\Delta t$ (e.g., 15 minutes) is denoted by $X_t \in \mathcal{R}^{|V| \times C}$, where $|V|$ denotes the number of sensors and $C$ is the dimensionality of node features (e.g., traffic volume and speed). 
\end{definition}

In the rest of the paper, we will use \emph{observation} place of \emph{spatio-temporal observation} when meaning is clear from the context.

\begin{definition}[Streaming Spatio-temporal Data Sequence]
Given a sensor network $G$ and a time period $\mathbb{T}_i$ containing $n$ consecutive time slots, i.e., $\mathbb{T}_i=\left\langle t_i^1, t_i^2,  \cdots, t_i^n\right\rangle$, a streaming spatio-temporal data sequence is a sequence of matrices, each representing an observation at a specific time slot $t_i^{j}$ with one sampling interval, 
where $t_i^{j+1}-t_i^{j}=\Delta t (1\le j\le n-1$).
In particular, a streaming spatio-temporal data sequence $D_i$ is a sequence of observations $D_i=\left\langle X_{t_i^1}, X_{t_i^2},  \cdots, X_{t_i^n}\right\rangle$, where $n$ is also the sequence length. %, $\Delta t$ denotes the sampling interval (e.g., 15min).
%In particular, the $i$-$th$ streaming spatio-temporal data $\mathcal{D}_i$ is a sequence of observations $\mathcal{D}_i=\left\langle X^{t_i-(j-1)\times \Delta t}_i, X^{t_{i}-(j-2)\times \Delta t}_i,  \cdots, X^{t_i}_i\right\rangle$ where $j$ is the sequence length, $\Delta t$ denotes the sampling interval (e.g., 15min).
%$k$-dimensional ($k\ge 1$) streaming spatio-temporal data is a sequence of $C$-dimensional observations $\mathcal{X}=\left\langle X^1, X^2,  \cdots, X^T\right\rangle$ with length $k$.
%where $T$ denotes the time interval (that is also the length of $\mathcal{X}$), and $X^t \in \mathcal{R}^{N \times C}$ denotes an $N \times C$ matrix that describes observations of all sensors at time $t$. Next, $N$ is the number of sensors and $C$ is the dimensions of node features (e.g., traffic volume, traffic speed).
%We define the network signal observations collected by all the sensors over $T$ time intervals as $\mathcal{X} = \{X^1, X^2,  \cdots, X^T\}$. The traffic observations at time slot $t$ is $X^t \in \mathcal{R}^{N \times C}$, where $C$ is the dimensions of node features (e.g., traffic volume, traffic speed). 
\end{definition}

%\begin{definition1}
%\textbf{Traditional Spatio-Temporal prediction (TSTF)} Given a road sensor network $G$, and historical observations over $M$ time intervals, traditional spatio-temporal prediction aims at learning a function $f(\cdot)$ to predict observations on next T time slots,
%\begin{equation}
%    [X^{t-M+1}, X^{t-M+2}, \cdots, X^{t}; G] \stackrel{f(\cdot)}{\longrightarrow} [\hat{X}^{t+1}, \hat{X}^{t+2}, \cdots, \hat{X}^{t+T}]
%\end{equation}
%where $X^t$ and $\hat{X}^{t}$ represent observation and prediction at time slot $t$, respectively.
%\end{definition1}

%\begin{definition}
%\textbf{Streaming spatio-temporal sets} Consider $m$ (probably infinite) streams of data $\mathcal{D} = \{D_1^{t_0:t_1}, D_2^{t_1:t_2}, \cdots, D_m^{t_{m-1}:t_{m}}\}$ arriving sequentially with the time evolving, each data $D_m$ contains a road sensor network $G$ and network signal observations $\mathcal{X}_i$ between time slots $t_{m-1}$ and $t_m$. We denote $D$ as streaming spatio-temporal sets, $D_1$ as a base set, and $D_2$ to $D_m$ as incremental sets.
%\end{definition}

Based on the above definitions, we formally define the  \textbf{Streaming Spatio-Temporal Prediction (SSTP)} problem as follows. %The main difference between TSTF and SSTF is that data are given sequentially in SSTF.

%\begin{definition1}\label{STCL_problem}
\textbf{SSTP Problem.}
Consider a sensor network $G$ that emits a sequence $\mathbb{D} = \left\langle D_1, D_2, \cdots,D_m \right\rangle$ ($m\ge 1$) of streaming spatio-temporal data sequences, where $D_i$ denotes a sequence of observations during time period $\mathbb{T}_i$ and $|D_i| =n$ (e.g., $\mathbb{T}_i$ is a day, and $n$ is $96$ when having $15$min interval). 
%contains spatio-temporal observations $\mathcal{X}_i$ between time slots $t_{m-1}$ and $t_m$. 
%\YAN{We denote %$\mathbb{D}$ as a streaming spatio-temporal set, $D_1$ as a base set and $D_2$ to $D_m$ as incremental sets.}
Given a current observation $X_{t_i^j}$ that is contained in $D_i$ ($1\le i\le m$), the SSTP problem aims to learn a function $f_i(\cdot)$ to predict $N$ future observations based on the current observation and its previous $M-1$ ($M < n$) historical observations while maximally preserving the learned knowledge from previous streaming data sequences $\left\langle D_1,\cdots, D_{i-1}\right\rangle$, i.e.,
%\begin{equation}
%    [X_{t_i^{j-M}}, \cdots, X_{t_i^j}; G] \stackrel{f_i(\cdot)}{\longrightarrow} %[\hat{X}_{t_i^{j+1}}, \cdots, \hat{X}_{t_i^{j+N}}]
%\end{equation}
\begin{equation}
\small
%\begin{split}
%\centering
[\overbrace{\cdots, X_{t_i^{j-1}}, X_{t_i^j}}^{M \text{ observations}}; G] \stackrel{f_i(\cdot)}{\longrightarrow}[\overbrace{X_{t_i^{j+1}}, X_{t_i^{j+2}, \cdots}}^{N \text{ observations}}],
%\end{split}
\end{equation}
where the learned knowledge from $\left\langle D_1,\cdots, D_{i-1}\right\rangle$ is preserved maximally.
%When dealing with a set $D_i$ ($1\le i\le m$), we aim to learn a function $f_i(\cdot)$ %for $D_i$ to map $M$ historical observations in $D_i$ ($M < n$) to $N$ future  observations, while maximally preserving the learned knowledge from previous streams of data $\{D_1,\cdots, D_{i-1}\}$, i.e.,
%Sequentially given a stream of data $D_m$ in streaming spatio-temporal sets, we aim to learn a function $f_m(\cdot)$ for $D_m$ to map $M$ historical observations ($M < t_m - t_{m-1}$) to $T$ future  observations, while maximally preserving the learned knowledge from previous streams of data $\{D_0,\cdots, D_{m-1}\}$, i.e., 
%\begin{equation}
%    [X^{t-(M-1)\times \Delta t}_m, \cdots, X^{t}_m; G] \stackrel{f_m(\cdot)}{\longrightarrow} [\hat{X}^{t+\Delta t}_m, \cdots, \hat{X}^{t+T\times \Delta t}_m]
%\end{equation}
%where $X_{t_i^j}$ represents an observation at time slot $t$ of $D_i$.
%\end{definition1}
%Noted that figure \ref{STCL} shows an illustration of replay-based streaming spatio-temporal data prediction.

\begin{figure*}[!h]
    \centering
    \includegraphics[scale=0.56]{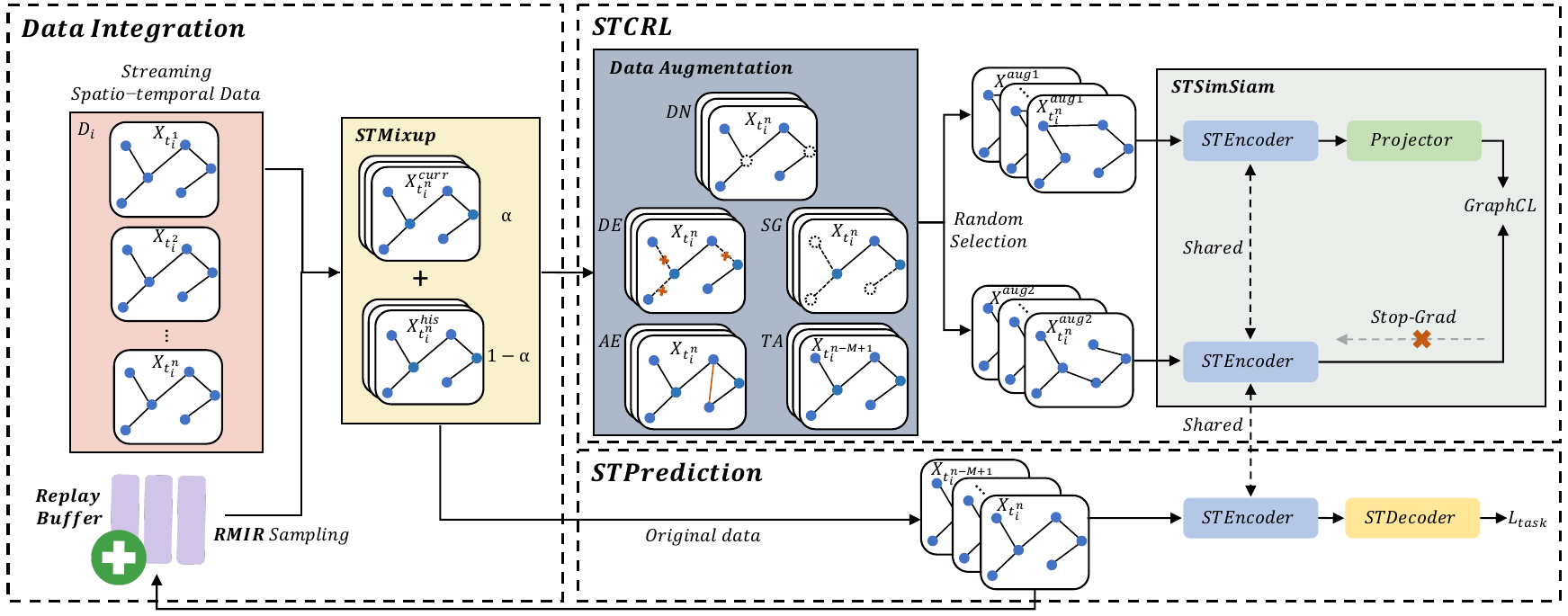}
    \vspace{-0.179cm}
    \caption{URCL Framework Overview}
    \vspace{-0.479cm}
    \label{overall_framework}
\end{figure*}

\section{Methodology}
\label{method}
We propose a framework, namely Unified Replay-based Continuous Learning (URCL), for spatio-temporal prediction on streaming data. We first give an overview of the framework and then provide specifics on each module in the framework.

\subsection{Framework Overview}
%Figure \ref{overall_framework} shows the framework of the proposed URCL. As shown in the figure, the model contains three major steps: data integration, spatio-temporal augmentation, continuous joint training and spatio-temporal data prediction. 

The framework consists of three major modules: data integration, spatio-temporal continuous representation learning (STCRL) and spatio-temporal prediction (STPrediction), as shown in Figure \ref{overall_framework}. 

% In the data integration step, given a stream of data $D_m$, as well as a replay buffer, we sample
\textbf{Data Integration.} Considering that we feed data into the framework in a stream, we first sample data from the current dataset $D_i$ and sample historical data from a replay buffer, which stores a subset of previously learned observations, by a ranking-based maximally interfered retrieval (RMIR) sampling strategy. Then we integrate the two kinds of data with the help of the proposed spatio-temporal mixup (STMixup) mechanism, to goal being to better accumulate the spatio-temporal knowledge and alleviate catastrophic forgetting.

\textbf{Spatio-Temporal Continuous Representation Learning (STCRL)}. In this module, a spatio-temporal SimSiam (STSimSiam) network, which is a variant of self-supervised Siamese networks~\cite{chen2021exploring}, is adopted for holistic representation learning by means of mutual information maximization. For the assistance of self-supervised learning, we propose five different data augmentation methods,  i.e., DropNodes (DN), DeleteEdges (DE), SubGraph (SG), AddEdge (AE), and TimeShifting (TS), based on the specific characteristics of spatio-temporal data. Then, the augmented data generated by two randomly selected augmentation methods are inserted into two spatio-temporal encoders (STEncoders) that share parameters with each other, followed by a projection MLP (projector). 

\textbf{Spatio-Temporal Prediction (STPredition).} We use an STEncoder that shares the same parameters with that in STCRL to learn latent hidden features from the original data, where the learned data are then stored in the replay buffer. Finally, the learned features are input into a spatio-temporal decoder (STDecoder) for prediction.

\subsection{Data Integration}
\label{data_intergration}
%As described in the problem statement, the streaming spatio-temporal data sequences are given sequentially with time evolving. 
To sample more representative samples from replay buffer $\mathcal{B}$, we design a novel ranking-based maximally interfered retrieval (RMIR) sampling method. Given the current data sequence $D_i$, we first sample M observations $\mathcal{X}_M = \left\langle X_{t_i^{k-M+1}}, \cdots X_{t_i^{k}} \right\rangle$ starting at time slot $t_i^{k-M+1}$ from $D_i$ and select samples $\mathcal{X}_{\mathcal{B}}$ stored in the replay buffer $\mathcal{B}$, where  $\mathcal{B}$ is designed to act as an explicit memory to maintain a subset of previously learned observations (i.e., the previously trained observations without STMixup). Then, the selected samples are combined with the current observations through a proposed spatio-temporal mixup (STMixup) mechanism to alleviate catastrophic forgetting by benefiting from historical observations. We can formulate the process of data integration as follows.
\begin{equation}
\small
\centering
\begin{split}
    \mathcal{X}_\mathcal{B} &= RMIR(\mathcal{B}, size=|\mathcal{S}|) \\
    \mathcal{X}_{mix} &= STMixup(\mathcal{X}_M, \mathcal{X}_\mathcal{B}),
\end{split}
\label{mix}
\end{equation}
where $|\mathcal{S}|$ is the sampling size.

We proceed to elaborate the \emph{RMIR sampling method} and the \emph{STMixup mechanism}.

\subsubsection{RMIR Sampling Method}
We design an RMIR sampling method to select $|\mathcal{S}|$ representative observations $\mathcal{X}_{\mathcal{B}}$ from the replay buffer. Typically, most methods in existing replay-based continuous learning methods randomly select observations from the replay memory, which leads to unsatisfied results about accuracy, due to the fact that they may ignore representative observations for replay. To select more representative samples, we first retrieve $|\mathcal{N}|$ observations $\mathcal{X}_{\mathcal{N}}$ which will be the most negatively impacted that suffer from an increase in loss by the update of foreseen parameters, where $|\mathcal{N}| \textgreater |\mathcal{S}|$. Specifically, given observation $X_{t_i^{\mathcal{B}}}$ in $\mathcal{B}$ and a standard objective function $\mathop{min}\limits_{\theta} \mathcal{L}_{RMIR}(f_\theta(X_{t_i^{\mathcal{B}}}), Y_{t_i^{\mathcal{B}}})$, where $f_{\theta}(\cdot)$ represents the URCL model, we update the parameters $\theta$ from $X_{t_i^{\mathcal{B}}}$ by gradient descent, as shown in Equation \ref{eq3}.
\begin{equation}
\centering
\label{eq3}
    \theta^v = \theta - \alpha\bigtriangledown\mathcal{L}_{RMIR}(f_\theta(X_{t_i^{\mathcal{B}}}), Y_{t_i^{\mathcal{B}}}),
\end{equation}
where $\mathcal{L}_{RMIR}$ denotes sampling loss, $f_\theta(\cdot)$ represents the model, and $Y_{t_i^{\mathcal{B}}}$ is the ground truth. Note that Mean Absolute Error (MAE) is adopted as sampling loss. We select the top-$\mathcal{N}$ values.% $\mathcal{X}_{\mathcal{N}}$ using the criterion $Top\mathcal{N}(\cdot)$ as follows.
% \begin{equation}
% \centering
% \begin{split}
%     \mathcal{X}_{\mathcal{N}} = Top\mathcal{N}(\mathop{Enum}\limits_{X_{t_i^{\mathcal{B}}} \in \mathcal{B}}(\mathcal{L}_{RMIR}(f_{\theta^v}((X_{t_i^{\mathcal{B}}}), Y_{t_i^{\mathcal{B}}}) \\- \mathcal{L}_{RMIR}(f_\theta((X_{t_i^{\mathcal{B}}}), Y_{t_i^{\mathcal{B}}})))),
% \end{split}
% \end{equation}
% where $\mathop{Enum}(\cdot)$ represents enumeration operation.

Considering temporal correlations (e.g., trend and periodicity) of spatio-temporal data, data from a long time ago (periodic data) have a significant impact on the current prediction due to similarities. We then calculate the similarities between observations in $\mathcal{X}_\mathcal{N}$ and $\mathcal{X}_M$ by Pearson coefficient. Finally, we sample top-$\mathcal{S}$ observations of $\mathcal{X}_\mathcal{N}$ that are most similar to $\mathcal{X}_M$. In this way, we can not only select samples that can alleviate catastrophic forgetting, but also select samples to enhance temporal dependency capturing.  

%Considering temporal correlations (e.g., trend, periodicity) of spatio-temporal data, data from a long time ago (period data) have significant impact on current prediction. But these data may rank lower in the top-$\mathcal{M}$ values
%that is not so negatively influenced by foreseen parameters
%, due that they are similar to current observations. If we simply select top $|\mathbf{B}|$ values from $X_{t_i^{\mathcal{B, M}}}$, the period data may be neglected, which will further impair the model's ability to capture temporal dependencies. To increase the possibility of these data being selected, we introduce randomness and stochastically select $|\mathbf{B}|$ instances $X_{t_i^{\mathcal{B}}}$ from $X_{t_i^{\mathcal{B, M}}}$ that are sampled. In this way, we can not only select samples that can alleviate catastrophic forgetting, but also select samples to enhance temporal dependencies capturing.  

\subsubsection{STMixup Mechanism}
To benefit from historical observations, we introduce STMixup mechanism to fuse current observations $\mathcal{X}_{M}$ and observations $\mathcal{X}_{\mathcal{B}}$ sampled in $\mathcal{B}$. In particular, STMixup interpolates between $\mathcal{X}_{M}$ and $\mathcal{X}_{\mathcal{B}}$ to encourage the model to behave linearly across streaming spatio-temporal data sequences to minimize catastrophic forgetting.

%Spatio-temporal mixup mechanism (STMixup) is proposed to fuse current instances and data sampled in $\mathcal{B}$ that is inspired by standard Mixup \cite{zhang2020does}. It can increment knowledge and encourage the model to behave linearly across a data stream.

Typically, assuming that $(x_i, y_i)$ and $(x_j, y_j)$ are two randomly selected feature-target pairs in the training data, STMixup generates virtual training examples by interpolation based on the principle of Vicinal Risk Minimization~\cite{chapelle2000vicinal} to enlarge the support of the training distribution that overcomes concept drift. 
In STMixup, we use observation-groundtruth pairs to represent feature-target pairs in training.
We use $(\Tilde{x}, \Tilde{y})$ to denote the interpolated feature-target pair in the vicinity of the raw two pairs.
\begin{equation}
\small
\begin{split}
    \Tilde{x} = \lambda \cdot x_i + (1 - \lambda) \cdot x_j \\
    \Tilde{y} = \lambda \cdot y_i + (1 - \lambda) \cdot y_j,
\end{split}
\end{equation}
where $\lambda \sim Beta(\alpha, \alpha)$, and $\alpha \in (0, \infty)$. We design STMixup by interpolating between current observations $\mathcal{X}_{M}$ and observations $\mathcal{X}_{\mathcal{B}}$ sampled from the replay buffer $\mathcal{B}$ by RMIR sampling. The interpolated observations $\mathcal{X}_{mix}$ after STMixup are formulated as follows:
\begin{equation}
\small
    \mathcal{X}_{mix} = \lambda \cdot \mathcal{X}_{M} + (1 - \lambda) \cdot \mathcal{X}_{\mathcal{B}}
\end{equation}
The interpolated observations $\mathcal{X}_{mix}$ can enhance a model's ability to learn continuously by revisiting past instances in the replay buffer $\mathcal{B}$, that would be most negatively impacted by foreseen parameters. Moreover, STMixup can introduce an approximation of a regularized loss minimization \cite{zhang2020does} to avoid overfitting.

\subsection{Spatio-temporal Continuous Representation Learning}
\label{STCRL}
The interpolated observations $\mathcal{X}_{mix}$ are input into the spatio-temporal continuous representation learning (STCRL) module for holistic feature learning. STCRL is a carefully designed self-supervised learning module, which consists of two parts: spatio-temporal data augmentation and an STSimsiam network. To enable effective spatio-temporal learning, we propose five customized spatio-temporal data augmentation methods by considering unique properties of spatio-temporal data, which transform a sample (i.e., observations in a sensor network) $\mathcal{G} = [\mathcal{X}_{mix}; G]$ to its corresponding perturbation $\mathcal{G}^{\prime}$. We randomly select two different perturbations $\mathcal{G}_1^{\prime}$ and $\mathcal{G}_2^{\prime}$, and then input them into an STSimSiam network, which consists of two STEncoders $f_{\theta_{STE}}$ and a projection MLP head $h(\cdot)$. The aim of STSimSiam is to better capture spatio-temporal dependencies. Finally, STSimSiam maximizes the mutual infomation between the representations of two selected perturbations to ensure holistic feature preservation.

We proceed to elaborate the \emph{spatio-temporal data augmentation} and the \emph{STSimSiam network}.

%The interpolated examples $\mathcal{X}_{mix}$ are input into the STSimSiam network for holistic feature learning under the guidance of self-supervised learning after data augmentation. The aim of employing self-supervised learning is making the model able to learn as many features as possible with mutual information maximization (e.g., InfoNCE). Aiming at assisting self-supervised learning, we propose five types of spatio-temporal augmentation methods by considering unique characteristics of spatio-temporal data to transform a sample $\mathcal{G} = [x^t_{mix}; G]$ to its correlated view $\mathcal{G}^{\prime}$.  Based on SimSiam\cite{chen2021exploring}, we further propose a STSimSiam network. STSimSiam consists of two STEncoders $f_{\theta_{STE}}$, which are composed of a spatio-temporal backbone network and are shared across a projection MLP and prediction MLP head $h(\cdot)$. In particular, STSimSiam maximizes the mutual information between the outputs of the projector and predictor MLP across two randomly selected different augmentations for $x^t_{mix}$.

\subsubsection{Spatio-Temporal Data Augmentation} We augment the interpolated observations $\mathcal{X}_{mix}$ using five carefully designed spatio-temporal augmentation methods, which builds semantically similar pairs and improves the quality of the learned representations by defeating perturbations. Although existing studies propose several data augmentation methods on graph data \cite{you2020graph, thakoor2021large, zeng2021contrastive}, they do not work well for spatio-temporal data due to complex spatial and temporal correlations, especially temporal correlations (e.g., closeness). Thus, we propose five spatially oriented data augmentation methods (i.e., DropNodes (DN), DropEdge (DE), SubGraph (SG), and AddEdge (AE)) and a temporally oriented method (i.e., TimeShifting (TS)). We cover each method next.

\begin{figure}
\centering
	\includegraphics[scale=0.8]{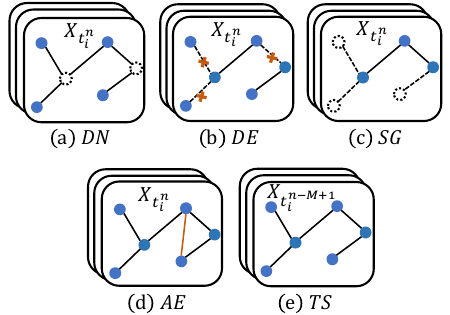}
	%\vspace{-0.3cm}
	\caption{Spatio-temporal Data Augmentation}
	\label{augment}
	\vspace{-0.5cm}
\end{figure}

\begin{itemize}
    \item \textbf{DN.} As shown in Figure~\ref{augment}(a), given a sample $\mathcal{G} = [\mathcal{X}_{mix}; G]$, DN randomly discards a certain proportion (e.g., 10\%) of the nodes in $G$ to get $\mathcal{G}^\prime = [\mathcal{X}_{mix}; G^\prime]$ following a definite distribution (e.g., a uniform distribution), which ensures that the missing nodes have no impact on the semantics (e.g., distribution) of $G$. In particular, we mask the entries in adjacency matrix $A$ of that correspond to the discarded nodes to pertube the graph structure.
    \begin{equation}
    \small
        A_{i, j}^\prime = \left\{
        \begin{array}{cl}
            0, & \mathrm{if}\, v_i\,\mathrm{is\,discarded}\\
            A_{i, j}, & \mathrm{otherwise},
        \end{array}
        \right.
    \end{equation}
    where $A_{i,j}^\prime$ is an entry of the augmented matrix $A^\prime$. Ideally, DN can promote model robustness by being less affected by missing data caused, e.g., sensor or communication failures. 
    
    %Given the graph $\mathcal{G}$, DN will randomly discard a certain proportion of nodes, which is motivated by node dropping in \cite{you2020graph}. We make the dropping possibility of each node follows a definite distribution (e.g., i.i.d. uniform distribution) to ensure that the semantic of $\mathcal{G}$ will not be affected by the missing vertices. Ideally, DN can promote the model robustness by defeating missing data caused by emergencies such as sensor failure, which is widely spread in spatio-temporal prediction tasks.
    \item \textbf{DE.} DE randomly drops a part of edges, as shown in Figure~\ref{augment}(b). As the weights of sensor network $G$ are important to describe the spatial correlations between nodes (e.g., distance and similarity). We first sample a certain ratio of edges $\mathcal{E}$ from $G$ following a specific distribution, and then set a threshold $\theta_{DE}$. If the weights of edges in $\mathcal{E}$ are lower than $\theta_{DE}$, we delete the corresponding edges, formulated as follows. 
    \begin{equation}
    \centering
    \small
        a_{i, j}^{'} = \left\{
        \begin{array}{cl}
            0, & \mathrm{if\, a_{i, j}\, < \,\theta_{DE}}  \\
            a_{i, j}, & \mathrm{otherwise},
        \end{array}
        \right.
    \end{equation}
    where $a_{i,j}$ denotes the weight of the edge between nodes $v_i$ and $v_j$, and $a_{i,j}^\prime$ is the updated weight. The aim of the threshold is to retain important connectives of edges. 
    
    %Edge deletion randomly drops a certain ratio of edges $\mathcal{E}_{D}$ on an unweighted graph followed a specific distribution. Nevertheless, weighted graphs are commonly used in spatio-temporal prediction tasks to quantify the local spatial correlations (i.e., the first law of Geography: \textit{"near things are more related than distant things"}) \cite{miao2022mba}. To accommodate with such case, we set a threshold $\theta_{DE}$ to avoid ignoring important connectives of edges. Specifically, if the weights of the randomly selected edges are lower than $\theta_{DE}$, we set the values of corresponding positions in adjacency matrix as 0. For example, assuming the edge between vertices $i$ and vertices $j$ is in $\mathcal{E}_{D}$, the corresponding weight $a_{i, j}$ will be changed by:
    %This method is not only applicable for predefined adjacency matrix that is mostly based on geographical distance \cite{li2018diffusion}, but also for adaptive adjacency matrix based on global spatial similarities\cite{bai2020adaptive}.
    \item \textbf{SG.} SG samples a subgraph $G^\prime = (V^\prime, E^\prime)$ from $G = (V, E)$ by random walk to maximally preserve the semantics of a sensor network (cf.\ Figure~\ref{augment}(c)), where $V^\prime \subseteq V$ and $E^\prime \subseteq E$. Through SG, we aim at improving local spatial correlations capturing by feature learning on the subgraph.
    \item \textbf{AE.} AE randomly selects a certain ratio of distant node pairs (e.g., more than three hops) and add edges between each node pair, which is shown in Figure~\ref{augment}(d). The weights of these added edges are set as the dot product similarities of 
    corresponding node pairs. Considering a node pair $(v_i, v_j)$, the corresponding weight $w_{i,j}$ is calculated as follows.
    \begin{equation}
    \small
        \begin{split}
            w_{i, j} = \vec{\mathcal{X}}^i_{mix} \cdot \vec{\mathcal{X}}^j_{mix},
        \end{split}
    \end{equation}
    where $\vec{\mathcal{X}}^i_{mix}$ is a vector that represents features of node $v_i$. The aim of AE is to strengthen the power of our model to capture global spatial correlations by connecting distant node pairs that are similar to each other.
    %The first law of geography may ignore global spatial correlations of spatio-temporal data in urban areas. For example, a commercial area may have few crowd flows coming from a park, although they are geographically close to each other, while a residential district far away may have a large number of people flowing into the commercial area. It is expected that AE is able to strengthen the model's ability to learn such global spatial dependencies.
    \item \textbf{TS.} As shown in Figure~\ref{augment}(e), TS, including time slicing, time warping, and time flipping, transforms current observations $\mathcal{X}_{mix}$ in $\mathcal{G}$ in the time domain. Note that we randomly select one of TS for model training.
    %Straightforwardly, TS augmentation methods transform spatio-temporal data in the time domain including time slicing, time warping and time flipping.
    \begin{enumerate}
        \item \textbf{\textit{Time Slicing.}} Time slicing sub-samples the current observations $\mathcal{X}_{mix}$ in the time domain by randomly extracting continuous slice $\mathcal{X}_{mix}^{slice}$ with length $l$. Formally,
        \begin{equation}
        \small
        \begin{split}
            \mathcal{X}_{mix}^{slice} = \left \langle X_{mix, t_i^{s-l+1}}, \cdots, X_{mix, t_i^{s}} \right \rangle,
        \end{split}
        \end{equation}
        where $t_i^{k-M+1} \leq t_i^{s-l+1} \leq t_i^{s} \leq t_i^{k}$, $t_i^k$ is current time slot, $M$ is the length of observations, and $X_{mix, t_i^{s}}$ denotes observations after STMixup at time slot $t_i^{s}$
        %Time slicing \cite{Wen0YSGWX21} is similar to cropping in CV area. It is a sub-sample method to randomly extract continuous slice in the time domain from the original spatio-temporal data.
        \item \textbf{\textit{Time Warping.}} Time warping upsamples sliced observations $\mathcal{X}_{mix}^{slice}$ by linear interpolation to generate warped observations $\mathcal{X}_{mix}^{warp}$, the length of it is equal to that of $\mathcal{X}_{mix}$.
        \begin{equation}
        \small
        \begin{split}
            \mathcal{X}_{mix}^{warp} = \left \langle X^\prime_{mix, t_i^{k-M+1}}, \cdots, X^\prime_{mix, t_i^{k}} \right \rangle,
        \end{split}
        \end{equation}
        where $X^\prime_{mix, t_i^{k}}$ represents the generated observations at time slot $t_i^{k}$.
        %Time warping selects a random time range of spatio-temporal data, and then compresses (down sample) or extends (up sample) it, while keeps other time range unchanged. Time warping would change the total time length of the original data, so it is always accompanied with time slicing for deep learning models \cite{muller2007dynamic}.
        \item \textbf{\textit{Time Flipping.}} Time flipping $\mathit{TF(\cdot)}$ flips the sign of warped observations $\mathcal{X}_{mix}^{warp}$ to generate a new sequence $\mathcal{X}_{mix}^{flip}$ in time domain as follows.
        \begin{equation}
        \small
        \begin{split}
            \mathcal{X}_{mix}^{flip} = \left \langle X^\prime_{mix, t_i^{k}}, \cdots, X^\prime_{mix, t_i^{k-M+1}} \right \rangle,
        \end{split}
        \end{equation}
        %Time flipping generate the new sequence $[x^{t}_{m}, x^{t-1}_{m}, \cdots, x^{t-M+1}_{m}]$ by flipping the sign of original data in $D_m$ $[x^{t-M+1}_{m}, x^{t-M+2}_{m}, \cdots, x^{t}_{m}]$ in time domain. 
    \end{enumerate}
\end{itemize}

We randomly apply two different data augmentation methods to the integrated observations $\mathcal{X}_{mix}$ (generated by STMixup) to obtain two augmented observations $\mathcal{X}_{mix}^{aug1}$ and $\mathcal{X}_{mix}^{aug2}$.

\subsubsection{STSimSiam Network}
Inspired by exciting feature learning ability of self-supervised learning, we design a novel STSimSiam network under the guidance of self-supervised learning to capture holistic spatio-temporal representations, inputting two randomly augmented observations $\mathcal{X}_{mix}^{aug1}$ and $\mathcal{X}_{mix}^{aug2}$. More specifically, STSimSiam consists of two STEncoders to learn spatio-temporal representations of $\mathcal{X}_{mix}^{aug1}$ and $\mathcal{X}_{mix}^{aug2}$, respectively, and a projection head to project latent embeddings of $\mathcal{X}_{mix}^{aug1}$ into the latent space of $\mathcal{X}_{mix}^{aug2}$. Finally, considering that mutual information maximization is proven to learn holistic features and thus improves continuous learning~\cite{guo2022online}, we maximize the mutual information between the representations learned from $\mathcal{X}_{mix}^{aug1}$ and $\mathcal{X}_{mix}^{aug2}$ by GraphCL loss.

As shown in the upper-right corner of Figure \ref{overall_framework}, we input two randomly augmented observations $\mathcal{X}_{mix}^{aug1}$ and $\mathcal{X}_{mix}^{aug2}$ into the STSimSiam network. We first encode the two augmented observations as two fixed vectors $z_1$ and $z_2$ by STEncoders $f_{\theta_E}$ which are made of a specific spatio-temporal network, i.e., GraphWaveNet in this work, to capture the complex spatio-temporal dependencies. Next, we feed $z_1$ into a projection MLP head to map it into the latent space of $z_2$, which contains several MLP layers denoted as $h(\cdot)$. The process can be formulated as follows.
\begin{equation}
    % \begin{split}
        z_1 = f_{\theta_E}(\mathcal{X}_{mix}^{aug1}),\quad
        p_1 = h(z_1),\quad
        z_2 = f_{\theta_E}(\mathcal{X}_{mix}^{aug2})
    % \end{split}
\end{equation}
We use stopgrad operation $SG(\cdot)$ \cite{madaan2021representational} to prevent the trivial solution obtained by STSimSiam. For example, $SG(z_{2})$ denotes that the STEncoder on $\mathcal{X}_{mix}^{aug2}$ receives no gradient from $z_{2}$.

To enhance holistic feature preservation, we employ mutual information maximization to maximize the similarities between representations of $\mathcal{X}_{mix}^{aug1}$ and $\mathcal{X}_{mix}^{aug2}$. First, for the current two augmented observations, we use cosine similarity to measure similarities $\mathcal{C(\cdot)}$ between their output vectors $p_1$ and $z_2$, formulated as follows.
\begin{equation}
\small
\begin{split}
    \mathcal{C}(p_{1}, z_{2}) &= \frac{p_{1}}{||p_{1}||_2}\cdot\frac{SG(z_{2})}{||SG(z_{2})||_2}\\
    & = \frac{h(f_{\theta}\mathcal{X}_{mix}^{aug1})}{||h(f_{\theta}(\mathcal{X}_{mix}^{aug1})||_2}\cdot\frac{SG(f_{\theta}(\mathcal{X}_{mix}^{aug2}))}{||SG(f_{\theta}(\mathcal{X}_{mix}^{aug2}))||_2},
\end{split}
\end{equation}
where $||\cdot||_2$ is $l_2$-$norm$.

We let $(\mathcal{X}_{mix}^{aug1}, \mathcal{X}_{mix}^{aug2})$ denote an augmented observation pair. For a minibatch of $\mathcal{S}$ augmented observation pairs, we adopt a GraphCL~\cite{you2020graph} loss \iffalse to minimize their negative mutual information \fi to maximize their mutual information. The GraphCL loss $L_{ssl}^s$ for the $s$-$th$ augmented observation pair is defined as follows.
\begin{equation}
\small
    \begin{split}
        L_{ssl}^s = -log\frac{exp(\mathcal{C}(p_{s,1}, z_{s,2})/\tau)}{\sum_{s^\prime=1, s^\prime\neq s}^{\mathcal{S}}exp(\mathcal{C}(p_{s,1}, z_{s^\prime,2})/\tau)},
    \end{split}
\label{ssl}
\end{equation}
where $p_{s,1}$ and $z_{s,2}$ denotes the output vectors of the $s$-$th$ augmented observation pair $(\mathcal{X}_{mix, s}^{aug1}, \mathcal{X}_{mix, s}^{aug2})$, $\mathcal{C}(\cdot)$ is cosine similarity, and $\tau$ is the temperature parameter. To extract more effective features~\cite{chen2021exploring}, we define a symmetric similarity function and obtain final $L_{ssl}^s$.
\begin{equation}
\small
%\label{graphcl}
    L_{ssl}^s = -log\frac{exp((\frac{1}{2}\mathcal{C}(p_{s,1}, z_{s,2})+\frac{1}{2}\mathcal{C}(p_{s,2}, z_{s,1}))/\tau)}{\sum_{s^\prime=1, s^\prime\neq s}^{\mathcal{S}}exp((\frac{1}{2}\mathcal{C}((p_{s,1}, z_{s^\prime,2})+\frac{1}{2}\mathcal{C}(p_{s,2}, z_{s^\prime,1}))/\tau)},
\label{ssl}
\end{equation}
where $p_{s,2}=h(f_{\theta_E}(\mathcal{X}_{mix, s}^{aug2}))$ and $z_{s, 1}=f_{\theta_E}(\mathcal{X}_{mix, s}^{aug1})$.

The final GraphCL loss is computed across all augmented pairs in the minibatch as follows.
\begin{equation}
\small
    L_{ssl} = \frac{1}{\mathcal{S}}\sum_{s=1}^{\mathcal{S}}L_{ssl}^s
\end{equation}
where $\mathcal{S}$ is the batch size.

%For a minibatch of $M$ examples are sampled and processed through STSimSiam network, GraphCL \cite{you2020graph} is considered in this paper which is a form of mutual information maximization between the latent representations of two kinds of augmented graphs and measured by dot product. The GraphCL loss for the $n$-$th$ sample is defined as: where $\tau$ represents the temperature parameter. The final GraphCL loss is computed across all augmented pairs in the minibatch.

\subsection{Spatio-Temporal Prediction}
\label{STDF}
We design a spatio-temporal prediction network including a spatio-temporal encoder (STEncoder) and a spatio-temporal decoder (STDecoder). In particular, we input the interpolated observations $\mathcal{X}_{mix}$ into STEncoder $f_{\theta_E}(\cdot)$ to learn high-dimensional representations through capturing complex spatio-temporal dependencies. The STEncoder shares parameters with the STEncoder in STSimSiam. The learned representations are input into STDecoder $f_{\theta_{D}}$ for prediction. Meanwhile, the recently learned data are stored in the replay buffer. The process can be formulated as follows.
\begin{equation}
\small
    % \begin{split}
        h_\theta = f_{\theta_E}(\mathcal{X}_{mix}), \quad
        \hat{\mathcal{Y}} = f_{\theta_{D}}(h_\theta),
    % \end{split}
\end{equation}
where $\hat{\mathcal{Y}}$ is the prediction.

\begin{figure}
    \centering
    \includegraphics[scale=0.8]{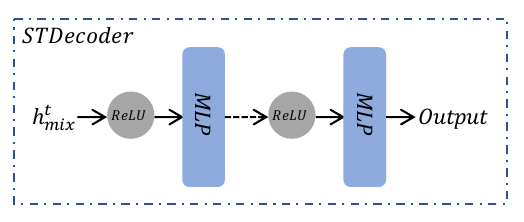}
    \vspace{-0.45cm}
    \caption{Illustation of STDecoder}
    \vspace{-0.5cm}
    \label{STDecoder}
\end{figure}

One of the advantages of our framework is its generality. It can easily serve as a plug-in for most existing spatio-temporal prediction models that follow the autoencoder architecture. GraphWaveNet \cite{wu2019graph} is one of the state-of-the-art spatio-temporal prediction models to capture precise spatial dependencies and long-term temporal dependencies, but which is not developed under the autoencoder architecture. Inspired by its excellent performance on deep spatio-temporal graph modeling, we take it as an example of spatio-temporal prediction models in our work to show how to reorganize its architecture so as to conform to the autoencoder architecture (i.e., STEncoder and STDecoder). Specifically, in STEncoder, graph convolution layers integrated with gated temporal convolution layers are employed to capture spatio-temporal dependencies, as shown in Figure \ref{STEncoder}, while several feed-forward networks are applied to map high-dimensional features into low-dimensional outputs for prediction in STDecoder, which is shown in Figure \ref{STDecoder}. It is particularly noticeable that we study the effect of different spatio-temporal prediction models (including RNN-based DCRNN~\cite{li2018diffusion} and attention-based GeoMAN~\cite{GeoMAN}) in our experimental part in Section 5.2.4. The studies show that our framework can adapt to different prediction models. For existing spatio-temporal prediction networks that lack an STDecoder, we employ stacked MLPs as the STDecoder.

%Generally, most existing models for spatio-temporal prediction can be divided into two-stages: spatio-temporal encoder (STEncoder) and spatio-temporal decoder (STDecoder), which are composed of backbones. One of the advantages of our framework is its generality which means that it is applicable for most spatio-temproal prediction models and can be a plug-in for them. In our work, the original mixed data $x_{mix}^t$ are input into the prediction network that contains a STEncoder and a STDecoder. The STEncoder maps data into a embedding latent space and captures the spatio-temporal dependencies of the data, while the learned features are then input into the STDecoder to decode the data representations for prediction. Moreover, we update the replay buffer by storing the most recent training samples and delete the oldest stored data.

\subsubsection{STEncoder}
\begin{figure}
    \centering
    \includegraphics[scale=0.65]{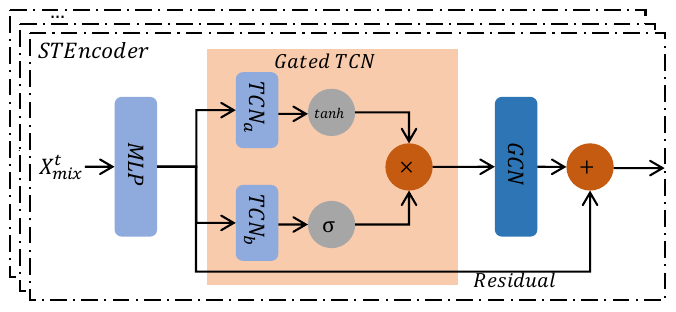}
    \vspace{-0.45cm}
    \caption{Illustation of STEncoder}
    \vspace{-0.4cm}
    \label{STEncoder}
\end{figure}
The architecture of STEncoder is shown in Figure \ref{STEncoder}. Taking $\mathcal{X}_{mix}$ as input, an MLP layer maps it into a high-dimensional latent space, and the learned features are then input into a Gated Temporal Convolution Network (TCN) layer to learn temporal correlations among observations of the input. Next, a Graph Convolutional Network (GCN) layer is used to capture spatial correlations among the observations, where a residual operation is used to ensure accuracy. The learned features of the $i$-$th$ spatio-temporal layer can be formulated as follows.
\begin{equation}
\small
    \begin{split}
        h_{\mathcal{\theta}}^i &= f_{\mathcal{G}}(\mathit{GatedTCN}(W_{\theta}^i\cdot h_{\theta}^{i-1} + b^i_{\theta}), A),
    \end{split}
\end{equation}
where $f_{\mathcal{G}}(\cdot)$ denotes GCN, $A$ is the adjacency matrix, $W_{\theta}^i$ is a learnable parameter, $b_{\theta}^i$ denotes the bias, and $h_{\theta}^0$ is euqal to $\mathcal{X}_{mix}$. The Gated TCN layer is composed of two parallel TCN layers (i.e., $\mathit{TCN_a}$ and $\mathit{TCN_b}$). 
%the STEncoder consists of stacked spatial-temporal layers each of which contains a graph convolutional layer (GCN) and a gated temporal convolution layer (Gated TCN) to capture complex spatial and temporal dependencies. 

\textbf{Graph Convolution Layer.} Recent studies pay considerable attention to generalize convolution networks for graph data. In this work, we use spectral convolutions on the constructed sensor network, which can be simply formulated as follows:
\begin{equation}
\small
\label{gcnformula}
    f_{\mathcal{G}}(X, A) = \sigma(\widetilde{A}XW^t),
\end{equation}
where $f_{\mathcal{G}}$ represents the GCN operation, $X$ denotes the node features, $\widetilde{A} = A + I_N$ is the adjacency matrix of $\mathcal{G}$ with added self-connections after normalizing, $W^t$ denotes the learnable weight matrix, and $\sigma(\cdot)$ is the activation function. 

Based on the first law of Geography: ''\textit{Near things are more related than distant things}''~\cite{zhang2017deep}, we first construct a local spatial graph by considering the geographical distance between nodes. If two nodes $v_i$ and $v_j$ are connected with each other geographically, an edge between them exists, and the corresponding weight is set as follows:
\begin{equation}
\small
        A_{i, j} = \left\{
        \begin{array}{cl}
            \frac{1}{dis}, & \text{if  $v_i$ connects to $v_j$}   \\
            0, & \text{otherwise},
        \end{array}
        \right.
    \end{equation}
where $dis$ denotes the geographical distance between two nodes $v_i$ and $v_j$. Following Diffusion Convolutional Recurrent Neural Network (DCRNN)~\cite{li2018diffusion} that adopts diffusion GCN, we generalize the diffusion convolution layer into the form of Equation \ref{gcnformula} by modeling the diffusion process of graph signals with $K$ finite steps, shown in Equation \ref{dgcn}.
\begin{equation}
\small
\label{dgcn}
    f_{\mathcal{G}}(X, A) = \sigma(\sum_{k=0}^KP_kXW_k),
\end{equation}
where $P_k$ denotes the power series of the transition matrix. For an undirected graph, we can get $P = \widetilde{A}/rowsum(\widetilde{A})$, while for a directed graph, there are two directions of the diffusion process: forward direction $P^f = \widetilde{A}/rowsum(\widetilde{A})$ and backward direction $P^b = \widetilde{A}^T/rowsum(\widetilde{A}^T)$. The diffusion graph convolution layer for a directed graph is derived as:
\begin{equation}
\small
    f_{\mathcal{G}}(X, A) = \sigma(\sum_{k=0}^KP_k^fXW_{k_1} + P_k^bXW_{k_2}),
\end{equation}

However, the first law of geography may not fully reflect the spatial correlation in urban areas, especially global spatial correlations (e.g., POI similarity)~\cite{yao2019revisiting, miao2022mba}. To tackle this problem, we construct a self-adaptive adjacency matrix $\widetilde{A}_{adp}$ by multiplying two randomly initialized node embeddings with learnable parameters $E_1$ and $E_2$:
\begin{equation}
\small
    \widetilde{A}^{adp} = \mathit{Softmax}(ReLU(E_1E_2^T)),
\end{equation}
where $\mathit{Softmax}$ is used to normalize the self-adaptive adjacency matrix. By comprehensively considering local and global spatial correlations, the final graph convolution layer is given in Equation~\ref{fgcn}.
\begin{equation}
\small
\label{fgcn}
    f_{\mathcal{G}}(X, A) = \sigma(\sum_{k=0}^KP_k^fXW_{k_1} + P_k^bXW_{k_2} + \widetilde{A}_k^{adp}XW_{k_3}),
\end{equation}
If the sensor network is unknown, we adopt the self-adaptive adjacency matrix alone to capture the spatial dependencies.

\textbf{Gated Convolution Layer.} To capture temporal dependencies, we employ the dilated causal convolution~\cite{yu2015multi} as our TCN due to its ability of modeling long-term temporal correlations and parallel computation. Specifically, given a data sequence $\mathcal{X}$ and a filter $f$, the dilated causal convolution operation of $x$ at $j$-th step is represented as:
\begin{equation}
\small
    \mathcal{X} \circ f(j) = \sum_{m=0}^{K-1} f(m)\mathcal{X}(j-d\times m),
\end{equation}
where $d$ denotes the dilation factor that reflects skipping steps, and $K$ is the length of filter $f$.

Gating mechanism is proven to be useful to control information flow through layers for TCN \cite{dauphin2017language}. We adopt a simple form of Gated TCN in our work to model complex temporal correlations, where the Gated TCN only consists of an output gate. Given the observations $\mathcal{X}$, the Gated TCN is illustrated in Equation~\ref{27}.
\begin{equation}
\small
\label{27}
    h = g(W_1 \times \mathcal{X} + b) \odot \sigma(W_2 \times \mathcal{X} + c),
\end{equation}
where $h$ is the learned features of input modeled by the first MLP layer, $W_1$ and $W_2$ are the learnable parameters, and $b, c$ are the bias, $\odot$ denotes the element-wise product, $g(\cdot)$ and $\sigma(\cdot)$ represent activation functions (e.g., tanh and sigmod).

\begin{algorithm}[t]
\caption{The URCL Framework}
\label{algorithm}
\begin{algorithmic}[1]
\Require Historical streaming spatio-temporal data sequences $\mathbb{D}$ from $D_1$ to $D_m$ and a sensor network $G$
\Ensure $i$-$th$ URCL model $\theta_i$
\For {$D_i$ in $\mathbb{D}$}
\State $D_i^{train}, D_i^{val}, D_i^{test} \longrightarrow \emptyset$
\label{line2}
% \For {$t \in \mathcal{T}$} 
%\label{line4}
\State put training, validating and testing instances into $D_i^{train}, D_i^{val}$ and  $D_i^{test}$, respectively
% \EndFor
\label{line5}
\While {$not$ $converge$}
\label{line6}
\State Sequentially select a batch of observations $D_i^{batch}$ from $D_i^{train}$
\State Sample previous observations $\mathcal{X}_{\mathcal{B}}$ from the replay buffer $\mathcal{B}$
\State $\mathcal{X}_{mix} \longleftarrow$ data fusion with STMixup by Equation \ref{mix}
\State $\mathcal{X}_{mix}^{aug1}, \mathcal{X}_{mix}^{aug2} \longleftarrow$ spatio-temporal data augmentation by $DN, DE, SG, AE$ and $TS$
\State $h_\theta \longleftarrow$ holistic feature learning by $STEncoder(\cdot)$ and GraphCL loss (i.e., Equation \ref{ssl})
\State $\hat{\mathcal{Y}} \longleftarrow$ prediction with $STDecoder(\cdot)$ by Equation \ref{STD}
\State Update $\theta_i$ based on Equation \ref{overalloss}
\EndWhile
\label{line14}
\State \textbf{Return} the learned $i$-$th$ learned model $\theta_i$
\EndFor
\end{algorithmic}
\end{algorithm}

\subsubsection{STDecoder}
The learned features are then input into the STDecoder to decode the data representations for prediction. As shown in Figure \ref{STDecoder}, the STDecoder contains several stacked feed-forward layers (i.e., MLPs) followed by activation functions (i.e., ReLU) to learn a projection function for future prediction. It can be formulated as follows.
\begin{equation}
\small
    \begin{split}
        \hat{\mathcal{Y}} = W_{\theta_D}(\alpha(h_\theta)+b_{\theta_D})),
    \end{split}
    \label{STD}
\end{equation}
where $\hat{\mathcal{Y}}$ denotes the prediction, $W_{\theta_D}$ is a learnable parameter, $\alpha(\cdot)$ is the ReLU activiation function, and $b_{\theta_D}$ is the bias.

\subsection{Overall Objective Function}
The overall objective of URCL is to minimize the prediction error by Mean Absolute Error (MAE) for each data stream $D_i$. The objective function is given in the following:
\begin{equation}
\small
    L_{task} = \frac{1}{\mathcal{L}} \sum_{l=1}^{\mathcal{L}}|\hat{\mathcal{Y}^l}-\mathcal{Y}^l|,
\end{equation}
where $\mathcal{L}$ is the training sample size, $\hat{\mathcal{Y}^l}$ is the prediction, and $\mathcal{Y}^l$ is the ground truth.

The final loss contains two parts: a prediction loss of the prediction task $L_{task}$ and a GraphCL loss $L_{ssl}$ for holistic feature learning. We combine them together and the overall loss is as follows.
\begin{equation}
\small
    L_{all} = L_{task} + L_{ssl}
    \label{overalloss}
\end{equation}
The whole process of the URCL is shown in Algorithm \ref{algorithm}, where lines $\ref{line2}$--$\ref{line5}$ state the data preprocessing and lines $\ref{line6}$--$\ref{line14}$ show the training process of URCL.

\section{Experimental Evaluation}
\label{experiment}
\begin{table}
\renewcommand\arraystretch{1}
	\small
	\centering
	\caption{\textit{Statistics of Datasets}}
	\vspace{-0.2cm}
	\setlength\tabcolsep{3.5pt}
	\scalebox{0.85}{
	\begin{tabular}{l|l|l|l|l}
		% \toprule[2pt]
		\hline
		\textbf{Dataset} & \textbf{\textit{METR-LA}} & \textbf{\textit{PEMS-BAY}} & \textbf{\textit{PEMS04}} & \textbf{\textit{PEMS08}} \\
		\hline
		%Data type & New York Bike Trip & New York Taxi Trip\\
		Area & \makecell[c]{Los\\Angeles} & California & \makecell[c]{San \\Francisco Bay} & \makecell[c]{San\\Bernaridino}\\\hline
		%Detectors & Unknown & Unknown & 3848 & 1979 \\\hline
		%\hline
		%Time span & 03/2012$\sim$ 06/2012& 01/2017 $\sim$ 05/2017 & 01/2018$\sim$02/2018 & 07/2016$\sim$08/2016 \\
		Time span & 4 months & 5 months & 2 months & 2 months \\\hline
		Sampling interval & 15 mins & 15 mins & 5 mins & 5 mins\\\hline
		No. of Nodes & 207 & 325 & 307 & 170 \\\hline
		Input steps & 12 & 12 & 12 & 12 \\\hline
		Output steps & 1 & 1 & 1 &1\\
		\hline
		% \bottomrule[2pt]
	\end{tabular}}
	\label{Dataset}
	\vspace{-0.3cm}
\end{table}
%In this section, we conduct extensive experimennts over four widely-used graph-based urban spatio-temporal datasets to evaluate the performance of the model. Noted that we choose traffic prediction on behalf of spatio-temporal prediction, as traffic prediction is one of the most popular spatio-temporal prediction tasks.
\begin{table*}
\renewcommand\arraystretch{1}
\small
    \centering
    \caption{\textit{Performance of Training on Streaming Data on Two Datasets}}
    \vspace{-0.2cm}
    \setlength\tabcolsep{3.4pt}
    \scalebox{0.9}{
    \begin{tabular}{c|c|c|c|c|c|c|c|c|c|c|c|c|c|c|c|c|c|c|c|c}
% \toprule[2pt]
\hline \multirow{3}{*} { \textbf{Method} } & \multicolumn{10}{c}{\textbf{PEMS-BAY}} & \multicolumn{10}{|c}{\textbf{PEMS08}} \\ \cline{2-21}& \multicolumn{5}{c|} { \textbf{MAE} } & \multicolumn{5}{c} { \textbf{RMSE} } & \multicolumn{5}{|c} { \textbf{MAE} } & \multicolumn{5}{|c} { \textbf{RMSE} } \\
\cline { 2 - 21 } & $\mathcal{B}_{set}$ & $\mathcal{I}_{set}^1$ & $\mathcal{I}_{set}^2$ & $\mathcal{I}_{set}^3$ & $\mathcal{I}_{set}^4$ & $\mathcal{B}_{set}$ & $\mathcal{I}_{set}^1$ & $\mathcal{I}_{set}^2$ & $\mathcal{I}_{set}^3$ & $\mathcal{I}_{set}^4$ & $\mathcal{B}_{set}$ & $\mathcal{I}_{set}^1$ & $\mathcal{I}_{set}^2$ & $\mathcal{I}_{set}^3$ & $\mathcal{I}_{set}^4$ & $\mathcal{B}_{set}$ & $\mathcal{I}_{set}^1$ & $\mathcal{I}_{set}^2$ & $\mathcal{I}_{set}^3$ & $\mathcal{I}_{set}^4$\\
\hline
\textbf{OneFitAll} & 1.21 & 2.76 & 2.72 & 3.75 & 3.63 & 2.17 & 5.82 & 4.88 & 7.28 & 6.86 & 18.15 & 38.75 & 39.32 & 44.22 & 27.21 & 26.28 & 44.29 & 57.34 & 63.01 & 37.19\\
%\hline
\textbf{FinetuneST} & 1.21 & 3.57 & 3.57 & 3.48 & 3.52 & 2.16 & 7.68 & 7.59 & 7.66 & 7.58 & 18.17 & 32.60 & 33.75 & 34.13 & 33.15 & 26.99 & 37.63 & 39.79 & 33.29 & 35.67\\
%\hline
\textbf{URCL} & \textbf{1.12} & \textbf{1.09} & \textbf{1.21} & \textbf{1.19} & \textbf{1.15} &\textbf{1.91} & \textbf{1.86} & \textbf{2.08} & \textbf{2.01} & \textbf{1.99} & \textbf{18.13} & \textbf{18.07} & \textbf{17.45} & \textbf{17.38} & \textbf{17.20} & \textbf{26.44} & \textbf{26.02} & \textbf{25.65} & \textbf{24.86} & \textbf{24.52}\\
\hline
%\hline
% \bottomrule[2pt]
\end{tabular}}
\label{cm_ofu}
\vspace{-0.4cm}
\end{table*}
\subsection{Experimental Setup}

\subsubsection{Datasets} 
The experiments are carried out on four widely-used graph-based urban spatio-temporal datasets: 
%We use the following commonly used four datasets for evaluation to enable fair comparison with existing studies and to facilitate reproducibility: 
\textit{METR-LA}, \textit{PEMS-BAY}, \textit{PEMS04} and \textit{PEMS08}, where \textit{METR-LA} and \textit{PEMS-BAY} are related to traffic speed prediction, and \textit{PEMS04} and \textit{PEMS08} are related to traffic flow prediction. 

\begin{itemize}
    \item \textbf{\textit{METR-LA.}} \textit{METR-LA} is collected in Los Angeles County and contains the traffic data from March to June 2012 as collected by 207 sensors. %The time span of this dataset is from March to June in 2012.

    \item \textbf{\textit{PEMS-BAY.}} \textit{PEMS-BAY} is collected in the Bay Area in California and contains data from 325 sensors. The time span is from January to May in 2017.

    \item \textbf{\textit{PEMS04.}} \textit{PEMS04} is collected from highways in the San Francisco Bay Area and contains 3848 detectors on 29 roads. The time span is from January to February in 2018.

    \item \textbf{\textit{PEMS08.}} \textit{PEMS08} is collectd in SanBernaridino from July to August in 2016. It contains the traffic data collected from 1,979 detectors on eight road segments.
\end{itemize}

We choose traffic prediction as a representative case of spatio-temporal prediction, as traffic prediction is a popular spatio-temporal prediction tasks. Dataset statistics are provided in Table~\ref{Dataset} and includes area, time span, sampling interval, number of nodes, and input and output steps.
%, and the detailed descriptions on the datasets are as follows.

\subsubsection{Evaluation Methods.}\label{sec:Evalumethod}
We compare URCL with the following baseline methods.
\begin{itemize}
    \item \textbf{\textit{ARIMA.}} The Auto-Regressive Integrated Moving Average (ARIMA) method is a classic statistic-based method for time series prediction~\cite{shekhar2007adaptive}.
    \item \textbf{\textit{DCRNN.}} The Diffusion Convolutional Recurrent Neural Network (DCRNN) method employs graph convolution networks to capture spatial correlations and the enocoder-decoder architecture with scheduled sampling to learn temporal correlations for traffic prediction~\cite{li2018diffusion}.
    \item \textbf{\textit{STGCN.}} The Spatio-Temporal Graph Convolutional Networks (STGCN) method applies ChebNet-GCN and 1D convolution to extract spatial and temporal dependencies for traffic prediction~\cite{yu2018spatio}.
    \item \textbf{\textit{MTGNN.}} The MTGNN model utilizes a graph-based deep learning method to exploit the inherent dependency relationships among multiple time series
    %a spatial GCN and GDCC to construct ST-blocks and applies stacked ST-blocks 
    for multivariate time series forecasting~\cite{wu2020connecting}.
    \item \textbf{\textit{AGCRN.}} The Adaptive Graph Convolutional Recurrent Network (AGCRN) model captures the fine-grained spatial and temporal correlations automatically in traffic series data for traffic prediction~\cite{bai2020adaptive}.
    \item \textbf{\textit{STGODE.}} The Spatial-Temporal Graph Ordinary Differential Equation Networks (STGODE) method adopts a tensor-based graph ordinary differential equation network to capture spatio-temporal dynamics and further is integrated with the temporal dilation convolution for traffic prediction~\cite{fang2021spatial}.
\end{itemize}

\begin{figure}
\centering
	\includegraphics[scale=0.6]{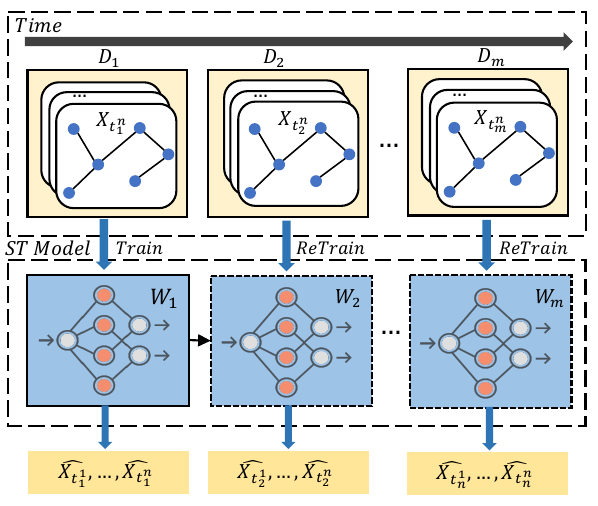}
	%\vspace{-0.3cm}
	\caption{Illustration of Replay-based Streaming Spatio-Temporal Data Prediction}
	\label{STCL}
	\vspace{-0.5cm}
\end{figure}

%To show the performance of URCL in terms of catastrophic forgetting, %and the necessities of the proposed replay-based framework, 
%we introduce the following compared methods.
%\begin{itemize}
 %   \item \textbf{\textit{OneFitAll.}} OneFitAll train a model with the base set, which is then used to predict all the testing data in the stream of spatio-temporal datasets.
%    \item \textbf{\textit{FinetuneST}} FinetuneST first trains an initial model on the base set, and then repeatedly fine-tunes the model with incremental sets.
%\end{itemize}
\begin{table*}[!htbp]
    \centering
    \renewcommand\arraystretch{0.9}
    \caption{Overall Accuracy on Four Datasets}
    % \vspace{-0.4cm}
    \setlength\tabcolsep{8.5pt}
    \begin{tabular}{c|c|c|ccccccc}
    
    % \toprule[2pt]
    \hline
        \multicolumn{2}{c|}{\textbf{Dataset}} & \textbf{Metric} & \textbf{ARIMA} & \textbf{DCRNN} & \textbf{STGCN} & \textbf{MTGNN} & \textbf{AGCRN} & \textbf{STGODE} & \textbf{URCL}  \\
        \hline
         \multirow{10}{*}{\textbf{METR-LA}} & \multirow{2}{*}{$\mathcal{B}_{set}$} & MAE & 4.99 & 4.01 & 3.92& 4.01& \underline{3.83}& 7.39& \textbf{3.82}\\
                                                                & & RMSE & 8.82& \underline{6.22} & 7.98 & 6.89 & 6.33 & 11.42 & \textbf{6.19}\\
         \cline{2-10} & \multirow{2}{*}{$\mathcal{I}_{set}^1$} & MAE & 5.87 & 10.55 & 9.99 & \underline{5.83} & 6.78 & 7.11 & \textbf{3.73}\\
                                                            & & RMSE & 9.77& 14.41& 14.77& \underline{9.71}& 12.42& 11.62& \textbf{6.14}\\
         \cline{2-10} & \multirow{2}{*}{$\mathcal{I}_{set}^2$} & MAE & 5.88 & 7.59 & 3.83& 4.33& \underline{4.30}& 7.35& \textbf{3.72}\\
                                                            & & RMSE & 10.49 & 8.44& \underline{5.63}& 6.89& 7.59& 11.77& \textbf{6.06}\\
         \cline{2-10} & \multirow{2}{*}{$\mathcal{I}_{set}^3$} & MAE & 4.88 & 4.77 & 4.96 & \underline{3.74}& 4.32& 6.45&\textbf{3.60}\\
                                                            & & RMSE & 10.49& 6.81 & 7.00& \underline{5.91}& 7.59& 10.17& \textbf{5.69}\\
         \cline{2-10} & \multirow{2}{*}{$\mathcal{I}_{set}^4$} & MAE & 5.00 & 4.13 & \underline{3.89}& 4.31& 4.32& 7.40& \textbf{3.74}\\
                                                            & & RMSE & 10.99 & 6.25& \textbf{5.63}& 6.94& 7.53& 11.78& \underline{6.01}\\
         
         \hline
         \multirow{10}{*}{\textbf{PEMS-BAY}} & \multirow{2}{*}{$\mathcal{B}_{set}$} & MAE & 1.98 & 1.30 & 1.31& 1.39& \underline{1.25}& 4.89& \textbf{1.12}\\
                                                                    & & RMSE & 3.85& \underline{2.24} & 2.28 & 2.53 & 2.51 & 7.93 & \textbf{1.92}\\
         \cline{2-10} & \multirow{2}{*}{$\mathcal{I}_{set}^1$} & MAE & 3.88 & 1.11 & 2.57 & 1.30 & \underline{1.14} & 4.66 & \textbf{1.09}\\
                                                                & & RMSE & 4.98& \underline{1.87}& 5.58& 2.22& 2.23& 7.61& \textbf{1.86}\\
         \cline{2-10} & \multirow{2}{*}{$\mathcal{I}_{set}^2$} & MAE & 3.67 & \underline{1.30} & 2.20& 1.47& 1.74& 2.44& \textbf{1.21}\\
                                                            & & RMSE & 4.89 & 2.34& 4.55& 2.53& \underline{2.24}& 3.69& \textbf{2.08}\\
         \cline{2-10} & \multirow{2}{*}{$\mathcal{I}_{set}^3$} & MAE & 3.24 & 1.21 & 3.33 & \underline{1.20}& 1.62& 5.31&\textbf{1.19}\\
                                                         & & RMSE & 6.77&2.11 & 7.06& 2.13& \underline{2.01}& 8.64& \textbf{2.00}\\
         \cline{2-10} & \multirow{2}{*}{$\mathcal{I}_{set}^4$} & MAE & 5.92 & 1.22 & 3.49& \underline{1.17}& 1.73& 4.84& \textbf{1.15}\\
                                                            & & RMSE & 10.22 & \underline{2.00}& 7.53& 2.10& 2.05& 7.84& \textbf{1.99}\\
         \hline
         \hline
         \multirow{10}{*}{\textbf{PEMS04}} & \multirow{2}{*}{$\mathcal{B}_{set}$} & MAE & 56.79 & 24.46 & 31.67& 22.99& \underline{22.09}& 36.54& \textbf{19.77}\\
                                                                    & & RMSE & 77.89& 33.85 & 45.33 & \underline{32.99} & 36.32 & 51.70 & \textbf{31.26}\\
         \cline{2-10} & \multirow{2}{*}{$\mathcal{I}_{set}^1$} & MAE & 65.78 & \underline{24.34} & 46.29 & 24.76 & 24.65 & 48.55 & \textbf{21.56}\\
                                                                & & RMSE & 87.94& 39.33& 62.39& \underline{35.33}& 40.22& 67.01& \textbf{34.10}\\
         \cline{2-10} & \multirow{2}{*}{$\mathcal{I}_{set}^2$} & MAE & 67.93 & 23.27 & 59.07& 29.70& \underline{21.93}& 49.60& \textbf{21.61}\\
                                                            & & RMSE & 79.02 & 36.68& 79.47& 39.64& \underline{35.97}& 67.40& \textbf{34.04}\\
         \cline{2-10} & \multirow{2}{*}{$\mathcal{I}_{set}^3$} & MAE & 63.92 & 25.95 & 29.08 & 26.04& \underline{21.34}& 49.99&\textbf{20.97}\\
                                                         & & RMSE & 75.02&35.28 & 33.26& 35.76& \underline{32.86}& 68.32& \textbf{32.12}\\
         \cline{2-10} & \multirow{2}{*}{$\mathcal{I}_{set}^4$} & MAE & 62.76 & 23.89 & 23.10& 25.81& \underline{20.81}& 49.81& \textbf{20.48}\\
                                                            & & RMSE & 78.22 & 33.42& 31.14& 35.79& \underline{32.66}& 67.69& \textbf{32.20}\\

         \hline
         \multirow{10}{*}{\textbf{PEMS08}} & \multirow{2}{*}{$\mathcal{B}_{set}$} & MAE & 45.98 & 23.07 & \underline{18.33}& 20.70& 23.03& 28.58& \textbf{18.08}\\
                                                                & & RMSE & 67.28& 32.24 & \underline{27.68} & 30.38 & 38.44 & 38.54 & \textbf{27.03}\\
         \cline{2-10} & \multirow{2}{*}{$\mathcal{I}_{set}^1$} & MAE & 55.22 & 20.35 & \underline{18.35} & 20.66 & 19.47 & 36.55 & \textbf{17.82}\\
                                                                & & RMSE & 75.33& 27.92& \textbf{24.30}& 29.82& 31.74& 46.04& \underline{26.37}\\
         \cline{2-10} & \multirow{2}{*}{$\mathcal{I}_{set}^2$} & MAE & 49.33 & 22.15 & 39.32& 20.27& \underline{17.45}& 36.06& \textbf{17.45}\\
                                                            & & RMSE & 59.89 & 29.63& 57.34& 28.97& \underline{27.35}& 45.25& \textbf{26.02}\\
         \cline{2-10} & \multirow{2}{*}{$\mathcal{I}_{set}^3$} & MAE & 58.99 & 19.94 & 44.22 & 22.68& \underline{17.70}& 36.12&\textbf{16.42}\\
                                                            & & RMSE & 65.79& \underline{27.04} & 63.01& 30.24& 27.56& 45.04& \textbf{24.88}\\
         \cline{2-10} & \multirow{2}{*}{$\mathcal{I}_{set}^4$} & MAE & 62.89 & 19.87 & 27.21& 18.92& \underline{18.35}& 36.05& \textbf{16.44}\\
                                                            & & RMSE & 71.36 & 27.91& 37.19& \underline{26.99}& 28.80& 45.30& \textbf{24.16}\\
         \hline
         % \bottomrule[2pt]
    \end{tabular}
    \label{baseline}
     \vspace{-0.45cm}
\end{table*}

\subsubsection{Metrics} Mean Absolute Error (MAE) and Root Mean Square Error (RMSE) are adopted as the evaluation metrics, which are defined as follows.
\begin{equation}
\small
    \begin{split}
        \mathit{MAE} &= \frac{1}{M}\sum^{M}_{m=1}|\hat{\mathcal{Y}^m}-\mathcal{Y}^m|\\
        \mathit{RMSE} &=  \sqrt{\frac{1}{M}\sum^{M}_{m=1}||\hat{\mathcal{Y}^m}-\mathcal{Y}^m||^2},
    \end{split}
\end{equation}
where $M$ is the testing data size, $\hat{y}^t$ is the prediction and $y^t$ represents the ground truth. %\YAN{Rewrite the equations and give mode detailed descriptions for these equations.}
The smaller the AME and the RMSE are, the more accurate the method is.
We also evaluate the efficiency of the models, including the training and inference (i.e., testing) time.

\subsubsection{Other Implementation Details} We implement our model with the Pytorch framework on NVIDIA Quadro RTX 8000 GPU. 

\textbf{Continuous Learning Settings.} We denote $D_1$ as a base set and $D_2$ to $D_m$ as incremental sets, where $\mathbb{D} = \left\langle D_1, D_2, \cdots, D_m \right\rangle$ ($m\ge 1$) is a sequence of streaming spatio-temporal data sequences as defined in Section~\ref{problem}.  
In the experiments, we use a base set, denoted as $\mathcal{\mathcal{B}}_{set}$, and four incremental sets, denoted as $\mathcal{I}_{set}^1$ to $\mathcal{I}_{set}^4$, to facilitate continuous training. Specifically, we use $30\%$ of each dataset as the base set and the remaining data in each dataset into four equal parts to form the incremental sets. 
The base set and incremental sets are provided sequentially with time evolving.
%\YAN{Why do you use $\mathcal{I}_{set}^1$ to $\mathcal{I}_{set}^4$ to denote $D_i$?}

\textbf{Training Process.}
We have tried to establish fair comparisons among the baseline methods (see Section~\ref{sec:Evalumethod}) by mapping their original training process into a continuous training processes, as shown in Figure~\ref{STCL}.
%the training process is as shown in Figure \ref{STCL}, where we 
%We first train an initial model on base set $D_1$ and then re-train a new model on incremental sets repeatedly.
We first train an initial spatio-temporal (ST) model on base set $D_1$ and then re-train a new ST model on incremental sets ($D_2$--$D_5$) based on the last learned model.

\textbf{Parameter Settings.} The model parameters are set as follows. The input sizes for \textit{METR-LA} and \textit{PEMS-BAY} are $12\times207\times2$ and $12\times325\times2$, respectively, where $12$ denotes the number of the previous time slots used for prediction, $307$ and $325$ are the numbers of the nodes, and $2$ is the number of channels representing traffic speed and flow. Moreover, the input size for \textit{PEMS04} and \textit{PEMS08} are $12\times307\times3$ and $12\times170\times3$, respectively, where $3$ denotes channels representing flow, speed, and occupancy. The STEncoder in URCL contains five layers, the hidden feature dimensions of which are 32, 32, 32, 32, and 256. The STDecoder contains two layers with 512 and 12 hidden features, respectively. The output of \textit{METR-LA}, \textit{PEMS-BAY}, \textit{PEMS04}, and \textit{PEMS08} are $1\times207\times1$, $1\times325\times1$, $1\times307\times1$, and $1\times170\times1$, respectively. We organize the buffer as a queue, whose size is set to 256. The parameters of the baseline methods are set based on their original papers and any accompanying code.  %or we use the publicly available code. 
%The parameters are set based on the original papers. 
%We provide the baseline parameter settings in the accompanying code repository \footnote[2]{\url{https://github.com/PaperCodeList/URCL}\label{code}}. 
Note that we normalize the streaming data into $[0, 1]$ to facilitate the feature learning.

\begin{figure*}[!htbp] \centering

	\subfigure[MAE on METR-LA] {
		\includegraphics[scale=0.258]{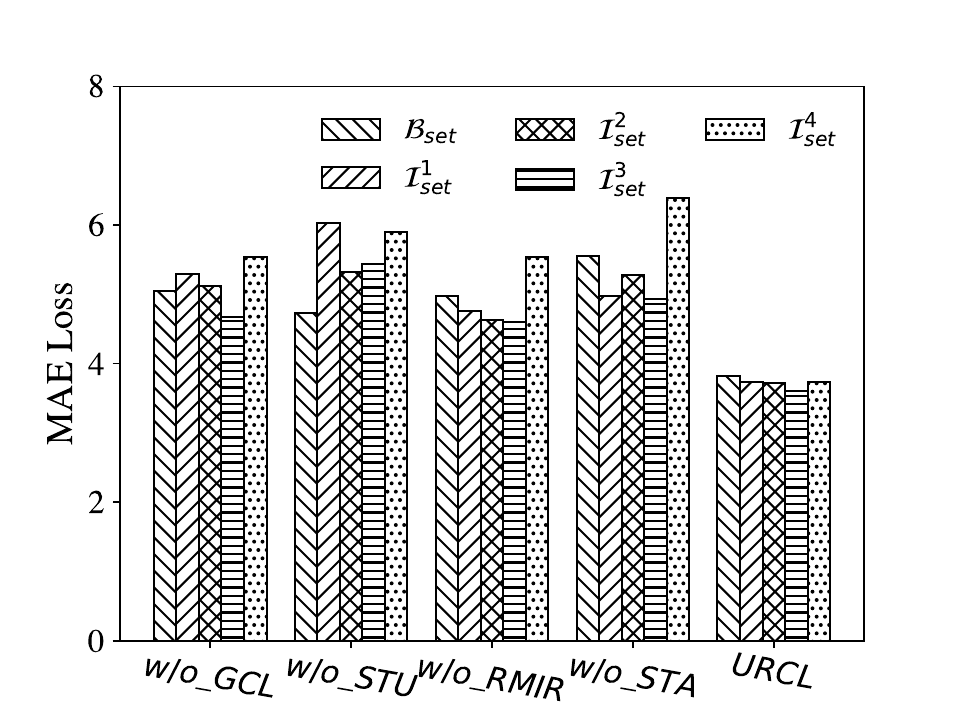} 
	}     
	\subfigure[RMSE on METR-LA] { 
		\includegraphics[scale=0.258]{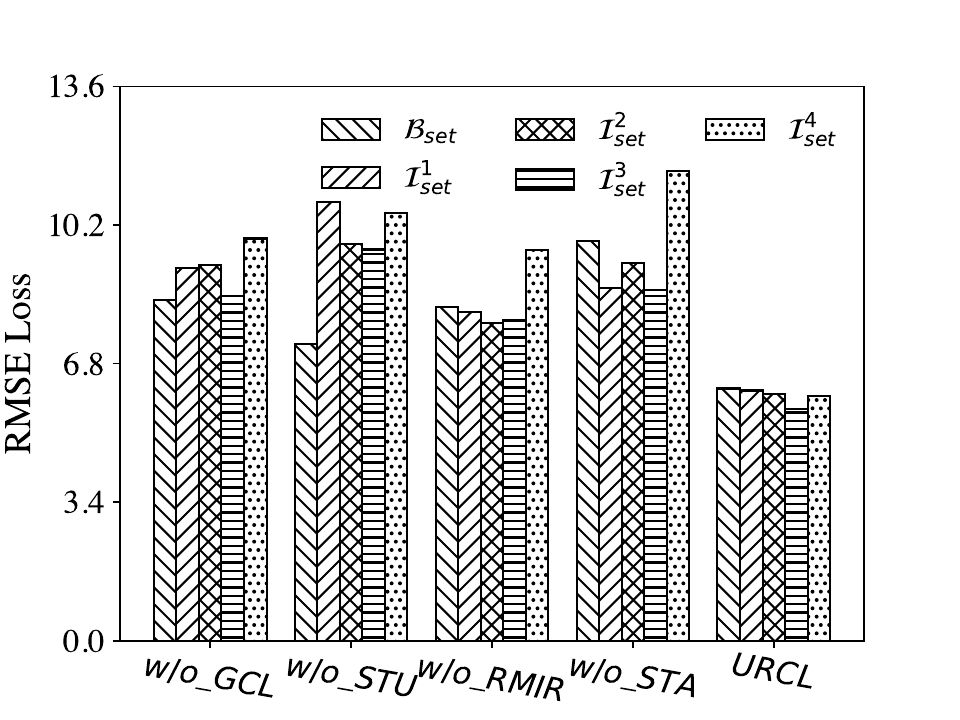}     
	}    
	\subfigure[MAE on PEMS08] { 
		\includegraphics[scale=0.258]{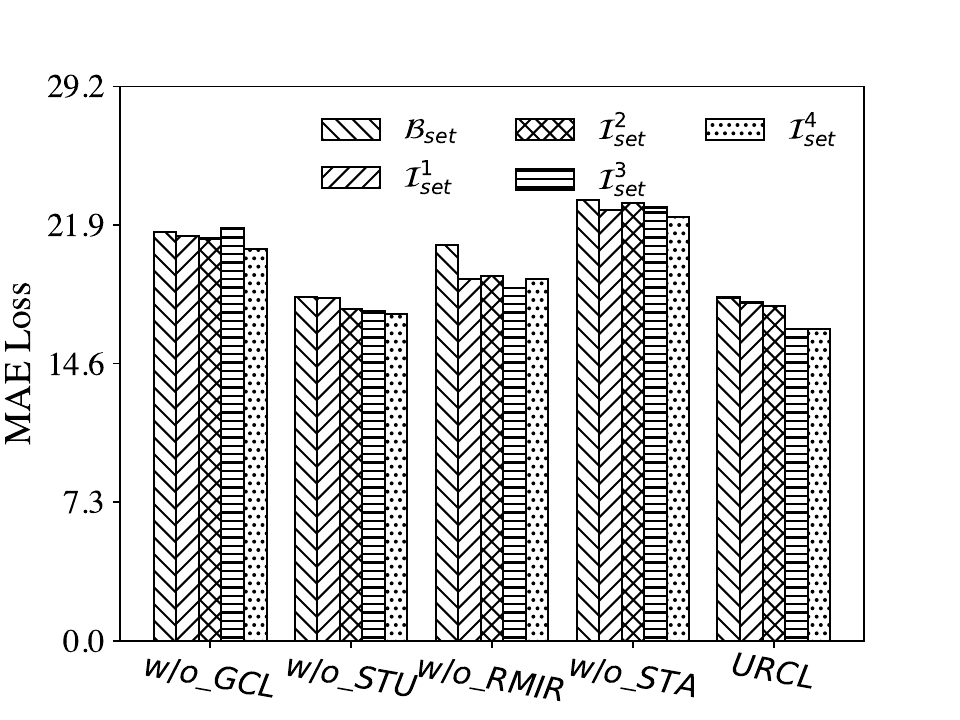}     
	}    
	\subfigure[RMSE on PEMS08] { 
		\includegraphics[scale=0.258]{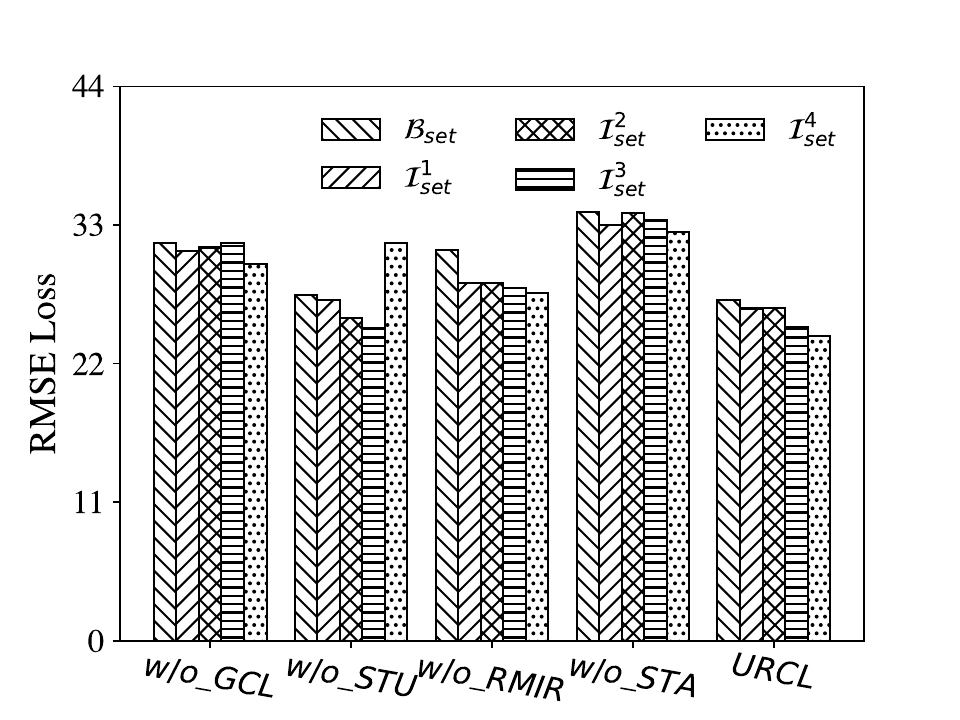}     
	}    
	\vspace{-0.3cm}
	\caption{RMSE and MAE of URCL and Its Variants}
	%\subfigure[NYCTaxi crowd flow]{
	%	\includegraphics[scale=0.1725]{image/taxi_traindata_Flow.pdf}}
	%\subfigure[NYCTaxi flow OD]{
	%	\includegraphics[scale=0.1725]{image/taxi_traindata_OD.pdf}}
	\vspace{-0.5cm}
	\label{ab_la} 
\end{figure*}

\subsection{Experimental Results}
%\subsection{Comparison with OneFitAll and FinetuneST}
\subsubsection{Performance of Training on Streaming Data}
%We first test whether there is a phenomenon of catastrophic forgetting of existing spatial-temporal prediction models
Our URCL proposes a replay-based strategy for training on streaming data, where a replay buffer is given to store previously learned samples that fuse with the training data by a spatio-temporal
mixup mechanism to preserve historical knowledge effectively.
%alleviating catastrophic forgetting.
To evaluate the performance of training on streaming data, two representative training strategies are used to replace the replay-based strategy in URCL and are compared with URCL: 1) \textbf{\textit{OneFitAll}} that trains a model with the base set and then predicts all the testing data in the stream of spatio-temporal datasets; and 2) \textbf{\textit{FinetuneST}} that trains an initial model on the base set and then fine-tunes the model with incremental sets repeatedly. GraphWaveNet is used as the base model of \textbf{\textit{OneFitAll}} and \textbf{\textit{FinetuneST}}.
%To validate our motivation about the phenomenon of catastrophic forgetting of existing spatio-temporal prediction models and the necessity of the proposed URCL model, we first compare the prediction performance with OneFitAll and FinetuneST in terms of MAE and RMSE. Note that we use a classic model STGCN \cite{yu2018spatio} as the netwrok of OneFitAll and FinetuneST. 
We report the MAE and RMSE results on the PEMS-BAY and the PEMS08 datasets in 
%The result is shown in 
Table~\ref{cm_ofu}, where the overall best performance is marked in bold. %\YAN{the best results are marked in bold.}
To save space, we do not report results on the METR-LA and PEMS04 datasets, as these are similar to those obtained for PEMS-BAY and the PEMS08.
The results on other datasets are included in the technical report\textsuperscript{\ref{code}}.

We can see that URCL shows the best performance in terms of MAE and RMSE when compared with other 
methods on both datasets. More specifically, URCL outperforms the best among the baselines by $14.5\%$--$67.3\%$ and $15.5\%$--$72.4\%$ for MAE and RMSE on PEMS-BAY, respectively, and $1.2\%$--$49.1\%$ and $8.9\%$--$35.5\%$ for MAE and RMSE on PEMS08, respectively. OneFitAll and FinetuneST offer acceptable MAE and RMSE results on the base sets on both datasets, but their performance deteriorates on incremental sets. OneFitAll shows that as time goes, the new data can be different from the training data, showing that concept drift happens and a static model does not work, calling for a CL model. While a simple CL based FinetuneST is not enough, as it has forgetting problems. 
URCL achieves relatively stable performance on both base sets and incremental sets demonstrating its superiority.  

\subsubsection{Overall Accuracy}
We report the MAE and RMSE values of the methods in Table~\ref{baseline}. % shows the comprehensive performance comparison among different methods over the four datasets. 
The best performance by an existing method (ARIMA, DCRNN, STGCN, MTGNN, AGCRN, and STGODE) is underlined, and the overall best performance is marked in bold.
%The best results are highlighted with bold font, and the best results achieved by baselines are underlined. 
Note that we repeatably train each original baseline on each base and incremental to enable 
%the streaming spatio-temproal datasets repeatedly 
fair comparison with the replay-based strategy (like Figure \ref{STCL}). 
The observations are in the following.

\begin{itemize}
    \item The proposed URCL achieves the best results among all baselines in most cases, performing better than the best among the baselines by up to $36.0\%$ and $34.1\%$ in terms of MAE and RMSE, respectively.
    In most cases, URCL outperforms the best among the baselines in terms of MAE and RMSE on the traffic speed datasets (i.e., METR-LA and PEMS-BAY) except for $\mathcal{I}_{set}^4$ on METR-LA. 
    Moreover, URCL also outperforms the best among the baselines in terms of MAE and RMSE on the traffic flow datasets (i.e., PEMS04 and PEMS08) except for $\mathcal{I}_{set}^1$ on PEMS08. 
    We observe that the performance improvements obtained by URCL on the traffic speed datasets exceed those on the traffic flow datasets. This is because because much more traffic flow data is available than traffic speed data. Thus, the baselines using the replay-based training strategy with sufficient data can get comparable results. %In addition, the results of Table \ref{baseline} verifies that URCL is more effective than existing state-of-the-art approaches on the four datasets.
    %\YAN{Check the results and reasons.}
 
    %with only two exceptions (5.63 RMSE achieved by STGCN over $\mathcal{I}_{set}^4$ of METR-LA and 24.30 RMSE achieved by STGCN over $\mathcal{I}_{set}^2$ of PEMS08).
    \item ARIMA, a traditional statistical method, performs worse than all methods except STGODE on METR-LA and PEMS-BAY and has the worst performance among all baselines on PEMS04 and PEMS08. This is 
    %among all the methods on all datasets. It is not surprising 
    because ARIMA only uses the time-series data of each region and ignores spatial dependencies.
    On METR-LA and PEMS-BAY, STGODE performs the worst since it is customized for speed prediction instead of flow prediction.
    \item As a popular spatio-temporal prediction model, AGCRN performs the best among all the baselines in most cases, due to its powerful ability to capture spatial and temporal correlations.

\end{itemize}
%One can see that the proposed URCL achieves the best results in most cases with only two exceptions (5.63 RMSE achieved by STGCN over $\mathcal{I}_{set}^4$ of METR-LA and 24.30 RMSE achieved by STGCN over $\mathcal{I}_{set}^2$ of PEMS08). 
%It shows that  
%As a popular spatio-temporal prediction model, AGCRN performs almost the best among all the baselines. URCL performs the best among all the methods, and outperforms state-of-the-art methods by a large margin on all datasets. 

%For PEMS-BAY, URCL reduces MAE by 10.4\% (from 1.25 to 1.12) on baseset $\mathcal{B}_{set}$ and 25.5\% (from 1.56 to 1.16) on incremental sets $\mathcal{I}_{set}^1$ to $\mathcal{I}_{set}^4$ compared with the best result achieved by the baseline AGCRN. For PEMS04, URCL obtains 10.5\%  MAE improvement (from 22.09 to 19.77) compared with the best result of baseline on $\mathcal{B}_{set}$ of PEMS04 and reduces RMSE averagely by 6.5\% (from 35.43 to 33.12) on incremental sets compared with the results achieved by baseline.

From the above observations, we can see that URCL can be applied on both traffic speed and traffic flow prediction tasks, which indicates generality across prediction tasks. %Table~\ref{baseline} shows that the performance improvements obtained by URCL on PEMS04 and PEMS08 are smaller than that on METR-LA and PEMS-BAY. It is reasonable because that traffic flow data is much more sufficient than traffic speed data and thus naively replaying baselines can get good results. In addition, the results of Table \ref{baseline} verifies that URCL is more effective than existing state-of-the-art approaches on the four datasets.

\subsubsection{Ablation Study}
To gain insight into the effects of key aspects of URCL, including the STMixup mechanism, the RMIR sampling strategy, the data augmentation methods, and the GraphCL loss, we evaluate four URCL
variants:
%We also compare the full version URCL with the following variants to further evaluate whether the key components used in our model are useful to the studied problem.
\begin{itemize}
    \item \textbf{\textit{w/o STMixup (w/o\_STU).}} \textit{w/o\_STU} is the URCL model without the STMixup module. We directly concatenate the original observations and sampled observations from the replay buffer. %Through comparing with it, we test whether the proposed STmixUp can help extract better features without forgetting historical knowledges and thus enhance the continuous learning performance.
    \item \textbf{\textit{w/o RMIR Sampling (w/o\_RMIR).}} \textit{w/o RMIR} replaces the RMIR sampling mechanism from URCL with a random sampling one. %Through comparing it, we test whether AM sampling can select most influential samples from replay buffer to enhance the model performance. 
    \item \textbf{\textit{w/o STAugmentation (w/o\_STA).}} \textit{w/o\_STA} is the URCL model without the STAugmentation. %Through comparing with it, we test whether the proposed augmentation methods is useful to the spatio-temporal continuous representation learning.
    \item \textbf{\textit{w/o GraphCL (w/o\_GCL).}} \textit{w/o\_GCL} is the URCL model without the GraphCL loss. %Through comparing with it, we test whether the GraphCL loss is useful for our model to learn the holistic representations.
\end{itemize}
%\YAN{Update the names of variants.}

%We give the results on the METR-LA and the PEMS08 datasets. %The results on the other two datasets are reported in the technical report\footnote{xxx}.
%To examine whether the proposed STMixup mechanism, AM sampling, spatio-temporal (ST) augmentation and GraphCL modules are all helpful to the studied problem, we perform the ablation study by comparing the full version of URCL with its variants w/o STU, w/o AMS, w/o STA and w/o GCL. The results are shown in 
Figure~\ref{ab_la} shows the MAE and RMSE results on METR-LA and PEMS08. 
Regardless of the datasets, URCL always outperforms its counterparts without the STMixup mechanism, the RMIR sampling strategy, the data augmentation methods, or the GraphCL loss.
These four components all improve the prediction accuracy of URCL, as removing any one of them increases the MAE and RMSE values on both base sets and incremental sets.
Further, on both datasets, \textit{w/o\_STA} performs the worst among all variants in most cases. URCL outperforms \textit{w/o\_STA}, reducing the MAE and RMSE values by up to $41.5\%$ and $47.8\%$, respectively, thus showing the benefit of spatio-temporal data augmentation. 
It is noteworthy that we include all the results in the technical report.

%the ST augmentation seems most significant because the prediction error increases remarkably when we remove ST augmentation module. Except augmentation methods, STMixup is more important on METR-LA dataset, while GraphCL is more important on PEMS08. Combining these modules together achieves the lowest RMSE and MAE, verifying that the four modules in URCL are effective for the studied problem.

\subsubsection{Effect of Different Backbones}
\begin{table}
\renewcommand\arraystretch{0.8}
\small
    \centering
    \caption{Effect of Various Backbones on METR-LA and PEMS04}
    \vspace{-0.2cm}
    \scalebox{0.85}{
    \begin{tabular}{ccccccc}
    %\toprule[2pt]
    % \toprule[2pt]
    \hline
        \multicolumn{2}{c}{\textbf{Dataset}} & \textbf{Metric} & \textbf{DCRNN} & \textbf{GeoMAN} & \textbf{URCL}  \\
        \hline
         \multirow{10}{*}{\textbf{METR-LA}} & \multirow{2}{*}{$\mathcal{B}_{set}$} & MAE & 4.97 & 5.11 & 3.82\\
                                                                & & RMSE & 8.18 & 8.47 & 6.19\\
         & \multirow{2}{*}{$\mathcal{I}_{set}^1$} & MAE & 4.76 & 4.87 & 3.73\\
                                                            & & RMSE & 8.07 & 8.28 & 6.14\\
         & \multirow{2}{*}{$\mathcal{I}_{set}^2$} & MAE & 4.64 & 4.41 & 3.72\\
                                                            & & RMSE & 7.80 & 7.09 & 6.06\\
         & \multirow{2}{*}{$\mathcal{I}_{set}^3$} & MAE & 4.61 & 4.39 & 3.60\\
                                                            & & RMSE & 7.86 & 7.28 & 5.69\\
         & \multirow{2}{*}{$\mathcal{I}_{set}^4$} & MAE & 5.54 & 5.27 & 3.74\\
                                                            & & RMSE & 9.59 & 8.77 & 6.01\\
                                                            
        \hline
         \multirow{10}{*}{\textbf{PEMS04}} & \multirow{2}{*}{$\mathcal{B}_{set}$} & MAE & 19.97& 19.78 & 19.77\\
                                                                & & RMSE & 32.25& 31.19 & 31.26\\
        & \multirow{2}{*}{$\mathcal{I}_{set}^1$} & MAE & 21.55 &20.82& 21.56\\
                                                            & & RMSE & 35.35 &33.15& 34.10\\
        & \multirow{2}{*}{$\mathcal{I}_{set}^2$} & MAE & 22.06 &21.07& 21.61\\
                                                            & & RMSE &35.92 &33.48 & 34.04\\
        & \multirow{2}{*}{$\mathcal{I}_{set}^3$} & MAE & 21.78 &21.49& 20.97\\
                                                            & & RMSE & 32.21 &34.34&32.12\\
        & \multirow{2}{*}{$\mathcal{I}_{set}^4$} & MAE & 21.24 &21.54&20.48\\
                                                            & & RMSE & 34.21 &34.75&32.20\\
         \hline
         % \bottomrule[2pt]
    \end{tabular}}
    \label{backbones}
    \vspace{-0.4cm}
\end{table}
Next, we study the effect of using different backbones, which are base models. For example, the backbone of URCL is the CNN-based GraphWaveNet~\cite{wu2020connecting} (see Section~\ref{STDF}).  %, where the overall best performance is marked in bold.
In addition to CNN-based models, two main streams of existing graph-based spatio-temporal prediction models exist, including RNN-based and attention-based models that adopt RNNs and the attention mechanism to learn temporal dynamics, respectively, and employ graph neural networks to capture spatial correlations. We select two representative models as the backbones of URCL, i.e., DCRNN~\cite{li2018diffusion} that is based on RNNs. GeoMAN~\cite{GeoMAN} that is based on the attention mechanism, %, and GraphWaveNet that is based on CNN and is our URCL method, 
We then compare them with URCL.
For simplicity, we use the names of the backbones as the names of the compared methods. 
%To show the ability of the proposed URCL to suit different backbones, we conduct experiments on two datasets for traffic speed prediction (METR-LA) and traffic flow prediction (PEMS04) by various classic backbones. 
%The results are shown in Table \ref{backbones}. Generally, there are three mainstreams of existing graph-based spatio-temporal prediction models: CNN-based methods, RNN-based methods and Attention-based methods that adopt CNN, RNN and Attention mechanism respectively to learn temporal dynamics, while employ graph neural network to capture spatial correlations. We select three classic and representative backbones DCRNN \cite{li2018diffusion} (RNN-based), GeoMAN \cite{GeoMAN} (Attention-based) and GraphWaveNet \cite{wu2020connecting} (CNN-based) from the three mainstreams.
Table~\ref{backbones} shows the prediction results on METR-LA and PEMS04. 
DCRNN and GeoMAN are neck-to-neck in terms of MAE and RMSE on both datasets.
URCL has the best performance in most cases, but the performance of the other two models are comparable especially on PEMS04, demonstrating the generality of URCL for adopting different backbones.
\begin{figure}[!htbp] \centering
	\vspace{-0.45cm}
	\subfigure[Training Time] {
		\includegraphics[scale=0.27]{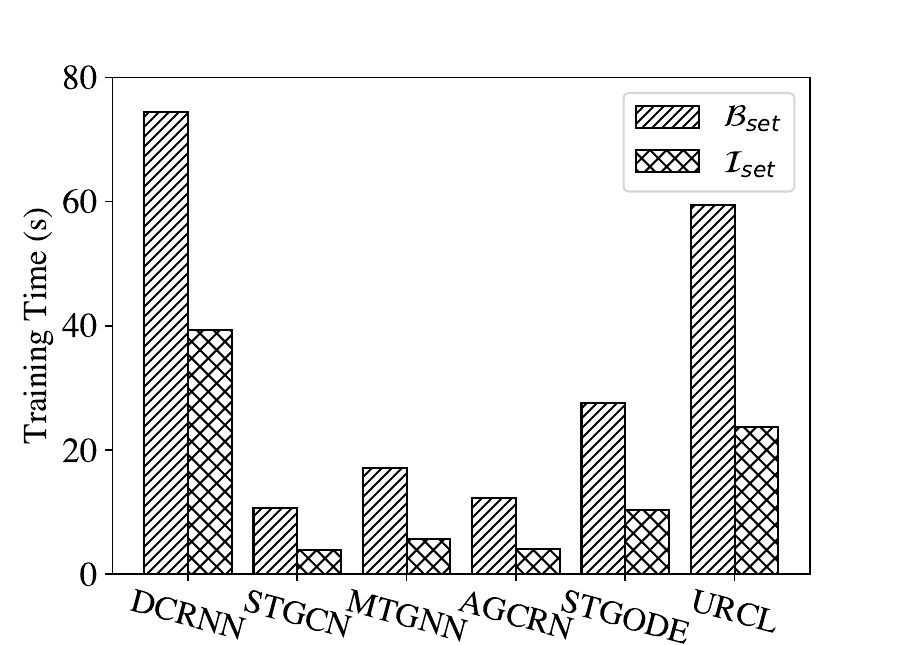} \label{trainingefficiency}
	}     
	\subfigure[Inference Time] { 
		\includegraphics[scale=0.27]{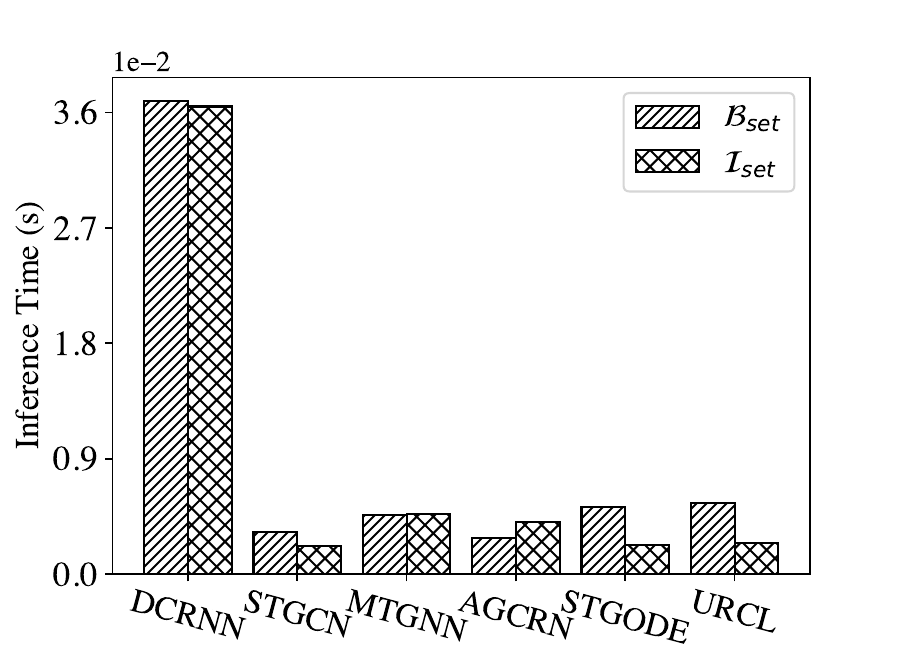}   \label{testingefficiency}  
	}  
	\vspace{-0.29cm}
	\caption{Training and Inference Time on PEMS04}
	%\subfigure[NYCTaxi crowd flow]{
	%	\includegraphics[scale=0.1725]{image/taxi_traindata_Flow.pdf}}
	%\subfigure[NYCTaxi flow OD]{
	%	\includegraphics[scale=0.1725]{image/taxi_traindata_OD.pdf}}
	\label{efficiency} 
	\vspace{-0.4cm}
\end{figure}

%we achieve comparable results by incorporating different backbones into URCL. It shows GraphWaveNet based URCL performs best, while the performance of DCRNN based URCL is inferior to GeoMAN based model. It is reasonable because DCRNN, GeoMAN and GraphWaveNet perform in descending order empirically for spatio-temporal prediction. The results show the generality of URCL to adapt various backbones.
\subsubsection{Efficiency}
As efficiency is important for spatio-temporal prediction on streaming data, we study the training time (of each epoch) and inference (testing) time (of each observation) for all the deep-learning-based methods on PEMS04. Figure~\ref{efficiency} shows the training and inference time of the base sets, as well as the averages of those of all the incremental sets. One can see that the training time of URCL is lower than that of DCRNN, as shown in Figure~\ref{trainingefficiency}. Although the baselines (except DCRNN) take less time for training, they perform worse than URCL in terms of prediction accuracy. Figure~\ref{testingefficiency} shows that the inference time of URCL is far lower than that of DCRNN and is comparable with those of the other baselines, which indicates the feasibility of URCL for model deployment in real streaming spatio-temporal prediction scenarios.

\subsubsection{Convergence Analysis of Training}
\begin{figure}[t]
\centering
	\includegraphics[scale=0.28]{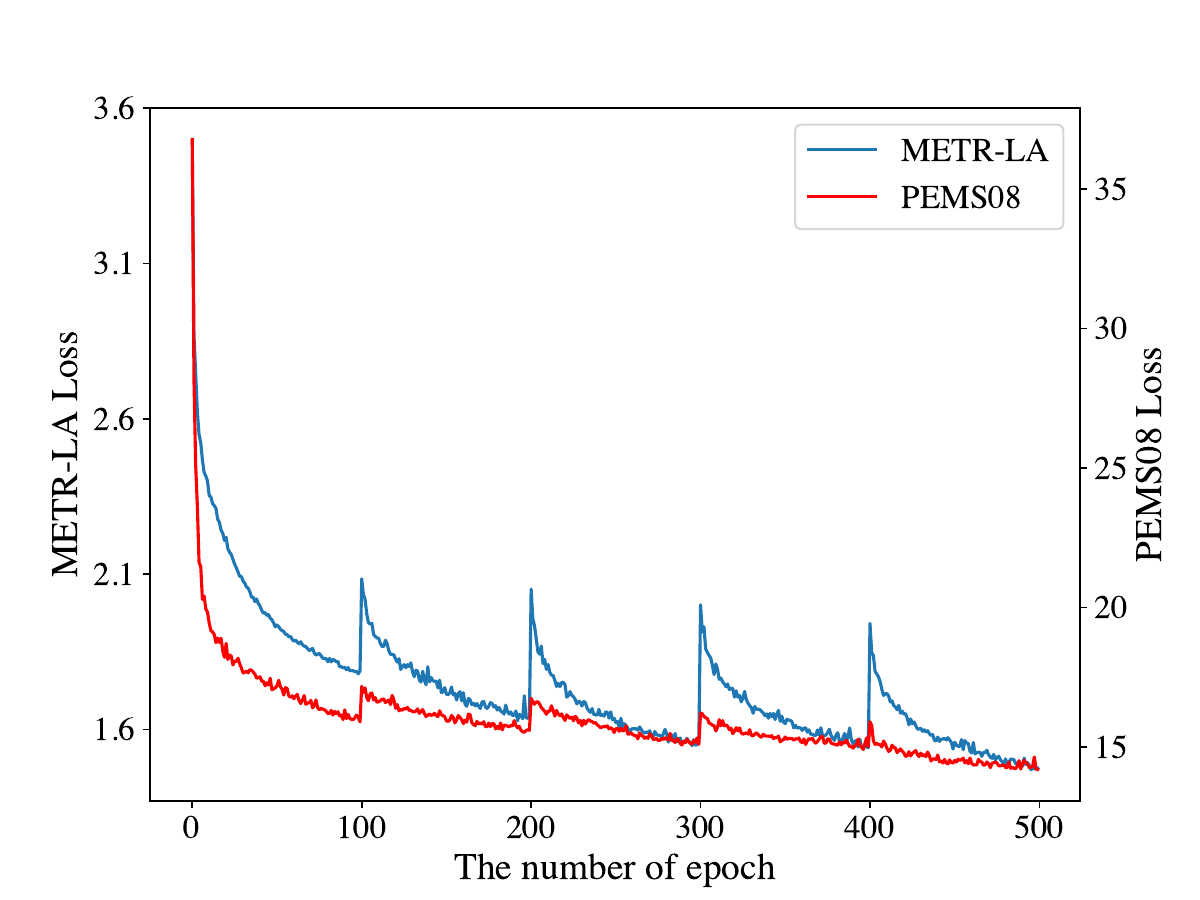}
 \vspace{-0.3cm}
	\caption{Training Convergence of URCL on MATR-LA and PEMS08}
	\label{loss}
	\vspace{-0.65cm}
\end{figure}

We study the training convergence of URCL on MATR-LA and PEMS08---see Figure~\ref{loss}. 
The number of epochs for training each base/incremental set is $100$, which means that we use the first $100$ epochs for training $\mathcal{B}_{set}$, the second $100$ epochs for training $\mathcal{I}^1_{set}$, the third $100$ epochs for training $\mathcal{I}^2_{set}$, and the rest can be done in the same manner.
%shows the training loss curve of the algorithm on the two datasets (METR-LA and PEMS08), where the 1-100, 101-200, 201-300, 301-400 and 401-500 epochs are the loss curve for the baseset $\mathcal{B}_{set}$,  and incremental sets $\mathcal{I}^1_{set}$, $\mathcal{I}^2_{set}$, $\mathcal{I}^3_{set}$, $\mathcal{I}^4_{set}$, respectively.
We observe that URCL converges after approximately $90$ epochs on $\mathcal{B}_{set}$ of both datasets, which shows that it converges quickly. Moreover, URCL converges after around $60$ epochs on all incremental sets of both datasets, meaning that it converges faster than on the base set.
This shows that URCL can reduce the training time on the unseen data substantially and thus can reduce computational costs. In addition, the training loss curves drop smoothly, after which there are minor fluctuations on the training loss. This is mainly because the proposed STMixUp mechanism  brings the benefits of regularization. %In the following experiment, we train URCL on all datasets 100 epochs.

%One can see the convergence speed of incremental set is quicker than that on baseset $\mathcal{B}_{set}$. It shows the proposed URCL can greatly reduce the training time on the unseen data and thus save computation cost. In addition, the loss curves drop smoothly, and there are still a few fluctuations on the loss during training. This mainly because the STMixUp mechanism we used which brings the benefits of regularization\cite{zhang2020does}. In the following experiment, we train URCL on all datasets 100 epochs.

\section{Conclusion}
\label{conclusion}
%This paper proposes a unified spatio-temporal continuous network named URCL for spatio-temporal prediction on streaming data to overcome catastrophic forgetting. By coupling a self-supervised learning module with spatio-temporal prediction network, the holistic features are effectively extracted through mutual information maximization. To embrace history knowledge, a STMixup mechanism is designed to integrate training data with representative data sampled in replay buffer. In addition, five spatio-temporal data augmentation methods is proposed to cope with the self-supervised learning module. Extensive evaluations on four real large datasets demonstrate the superior performance of the proposed model on various spatio-temporal prediction tasks. In the future, it would be also interesting to further apply URCL to the task of POI recommendation.

We propose a unified replay-based continuous learning framework for spatio-temporal prediction on streaming data, that aims to capture the complex patterns in spatio-temporal data.  To embrace historical knowledge, an STMixup mechanism is designed to integrate training data with representative data sampled from a replay buffer. By coupling a spatio-temporal SimSiam network with a spatio-temporal prediction autoencoder, the holistic features are extracted through mutual information maximization. In addition, five spatio-temporal data augmentation methods are proposed to cope with the SimSiam network. An empirical study with real datasets offers evidence that the paper’s proposals improve on the state of the art in terms of prediction accuracy. An interesting research direction is to attempt to further improve the training efficiency of the proposed URCL.

\section{Acknowledgment}
This work is partially supported by Independent Research Fund Denmark under agreements 8022-00246B and 8048-00038B, the VILLUM FONDEN under agreements 34328 and 40567, Huawei Cloud Database Innovation Lab, and the Innovation Fund Denmark center, DIREC.

%\clearpage

\bibliographystyle{IEEEtran}
\bibliography{sample}

\end{document}